\documentclass[]{aa}  

\usepackage{graphicx}
\usepackage{txfonts}
\usepackage[]{hyperref}
\begin{document} 

    \title{Measurement of the open magnetic flux in the inner heliosphere down to 0.13AU}

    \author{Samuel T. Badman \inst{1,2}
            \and
            Stuart D. Bale \inst{1,2,3}
            \and 
            Alexis P. Rouillard \inst{4}
            \and
            Trevor A. Bowen \inst{2}
            \and 
            John W. Bonnell \inst{2}
            \and
            Keith Goetz \inst{5}
            \and 
            Peter R Harvey \inst{2}
            \and 
            Robert J. MacDowall \inst{6}
            \and 
            David M. Malaspina \inst{7}
            \and
            Marc Pulupa \inst{2}
    }
    
    \institute{Physics Department, University of California, Berkeley, CA 94720-7300, USA
                \email{samuel\_badman@berkeley.edu}
                \and Space Sciences Laboratory, University of California, Berkeley, CA 94720-7450, USA
                \and The Blackett Laboratory, Imperial College London, London, SW7 2AZ, UK
                \and IRAP, Université Toulouse III - Paul Sabatier, CNRS, CNES, Toulouse, France
                \and School of Physics and Astronomy, University of Minnesota,
        Minneapolis, MN 55455, USA
                \and Solar System Exploration Division, NASA/Goddard Space Flight Center, Greenbelt, MD, 20771
                \and Laboratory for Atmospheric and Space Physics, University of Colorado, Boulder, CO 80303, USA
    }

   \date{Submitted to A\&A September 11 2020; Revised November 1 2020; Accepted November 4 2020}

  \abstract
   {Robustly interpreting sets of in situ spacecraft data of the heliospheric magnetic field (HMF) for the purpose of probing the total unsigned magnetic flux in the heliosphere is critical for constraining global coronal models as well as understanding the large scale structure of the heliosphere itself. The heliospheric flux ($\Phi_H$) is expected to be a spatially conserved quantity with a possible secular dependence on the solar cycle and equal to the measured radial component of the HMF weighted by the square of the measurement's heliographic distance ($B_R R^2$). It is also expected to constitute a direct measurement of the total unsigned magnetic flux escaping the corona ($\Phi_{open}$). Previous work indicates that measurements of $\Phi_H$ exceed the value predicted by standard coronal models (the `open flux problem'). However, the value of the open flux derived from in situ measurements remains uncertain because it depends on the method employed to derive it. Past derivations also pointed towards an increase in $\Phi_H$ with heliocentric distance, although this may also be related to its method of computation.}
   {In this work, we attempt to determine a more robust estimate of the heliospheric magnetic flux ($\Phi_H$) using data from the FIELDS instrument on board Parker Solar Probe (PSP) , to analyse how susceptible it is to overestimation and a dependence on time and space, as well as considering how it compares to simple estimates of $\Phi_{open}$ from potential field source surface (PFSS) models.}
   {We compared computations of the heliospheric magnetic flux using different methods of data processing on magnetic field data from PSP, STEREO A, and Wind. Measured radial trends in fluctuations and background magnetic structure were used to generate synthetic data to analyse their effect on the estimate of $B_R R^2$. The resulting best estimates were computed as a function of time and space and then compared to estimates from PFSS models. }
  { Radially varying fluctuations of the HMF vector as well as large-scale variations in the inclination of the Parker spiral angle are shown to have a non-trivial effect on the 1D distributions of $B_R R^2$. This causes the standard statistical metrics of the mean and mode (the most probable values) to evolve with radius, independently of the central value about which the vector fluctuates. In particular, the mean systematically underestimates $\Phi_H$ for R < 0.8AU and increases close to 1AU. We attempt to mitigate for this by using the `Parker spiral method' of projecting the vector onto the background Parker spiral direction (which requires vector fluctuations to be evenly distributed about a central value). Even with this method, we find evidence of a small enhancement in flux close to 1AU. The fraction of field which is locally inverted in a given time interval grows with radial distance from the Sun which remains a possible physical reason for this excess but is essentially negligible at PSP’s perihelia distances where the impact of fluctuations in general is also much reduced. The Parker spiral method (PSM) and most probable values converge close to the Sun. Our derived best estimate for the time interval studied is $\sim2.5^{+0.3}_{-0.6}$ nT AU$^2$. To the extent probed by PSP, no strong dependence on latitude or longitude is apparent, although at 1AU, the spread of measured values appears to grow at the highest latitudes. The best estimate of the heliospheric flux is significantly larger than estimates from PFSS models studied here, which predict values from 1.2-1.8 nT AU$^2$, depending on the choice of magnetogram or source surface height.}
   {Of the methods  for computing the heliospheric flux over a wide range of heliocentric distances using only magnetic field data considered in this work, the most robust choice is to use the PSM. The decay of fluctuations and weakening importance of local flux inversions at smaller heliocentric distances indicate that the measurement is most accurate close to the sun and that it is justified for us to consider that $\Phi_H \sim \Phi_{open}$ for these measurements. The determined value is too high to be explained via PFSS models. Contemporary magnetohydrodynamic (MHD) models with the same photospheric input are unlikely to close this gap. Therefore, the most likely solutions remain in improvements of coronal models, for example, through improved boundary conditions via the direct measurement of the photospheric field in the solar polar regions or through the inclusion of missing physical processes such as time-dependent or non-potential effects, which can produce a contribution to the open flux that is not rooted in obvious coronal holes.}    
   \maketitle

   \clearpage
%
\section{Introduction}
\label{sec:Introduction}

The coronal magnetic field may be topologically separated into closed field lines, which form loops on coronal length scales that confine coronal plasma, and open field lines, where the dynamic pressure of the out-flowing solar wind dominates over magnetic stresses, thus causing the magnetic field lines to be advected outwards into the heliosphere. Over interplanetary length scales, the solar rotation and radially out-flowing solar wind combine to form the well-established Parker spiral magnetic field \citep{Parker1958}. The total unsigned open magnetic flux is the total flux ($|\boldsymbol{B}\cdot d\boldsymbol{S}|$) carried by each open coronal field line, summed over any closed surface encompassing the Sun, but usually integrated over a spherical heliocentric surface. Here, $\boldsymbol{B}$ denotes the coronal or heliospheric magnetic field (HMF) vector, and $d\boldsymbol{S}$ is the differential surface element that the field line intersects. We note the magnitude is taken since the signed open flux cancels out over a closed surface integral, according to Gauss' Law ($\nabla \cdot \boldsymbol{B} = 0$).  In this paper, we refer to this integrated quantity as the `open flux', that is, $\Phi_{open}$.

The open flux is a quantity of significant interest in coronal and heliospheric physics. It determines the HMF field strength, which, in turn, affects the coupling of the solar wind with planetary magnetospheres. It determines the transport properties of cosmic rays through the heliosphere \citep{Cliver2013}. It has been shown to vary with solar cycle \citep[e.g.  ][]{Wang2000} and, therefore, it may carry information about the sun's internal dynamo. Finally, given the expected conservation of the quantity, direct measurements of the open flux in interplanetary space, which is the main subject of this paper, can be used to constrain global coronal models for which the open flux is an observable. Typically, interplanetary measurements of $\Phi_{open}$ exceed the estimates coming from most global coronal models \citep[the `open flux problem', ][]{Linker2017}, with the agreement worsening at solar maximum \citep{Wallace2019}.

As hinted above, there are two typical contexts in which the open flux is computed. The first is via global coronal models. Such models take maps of the photospheric magnetic field obtained by remote measurements of Zeeman splitting of a photospheric emission line and utilise these boundary conditions to extrapolate a 3D coronal field. The two most common types of global models are the potential field source surface extrapolation \citep[PFSS][]{Altschuler1969,Schatten1969,Wang1992} and magnetohydrodynamics \citep[MHD, e.g.  ][]{Lionello2008}. 

 In PFSS models, the outer boundary of the model (called the source surface) is a sphere at a fixed radius ($R_{SS}$) at which all intersecting field lines are defined to be radial and open to the solar wind. In MHD models, there is no explicit outer boundary and instead, field lines may be traced and a `source surface' determined numerically by mapping the region where field lines become radial. The open flux is then computed by integrating the modelled magnetic field $\boldsymbol{B}(R,\theta,\phi)$ over this outer boundary (we note that $\theta$ and $\phi$ here refer to heliographic latitude and longitude, respectively):

\begin{align}
    \Phi_{open} = \int_0^{2\pi} \int_{-\frac{\pi}{2}}^{\frac{\pi}{2}} |B_R(\theta,\phi,R=R_{SS})| R_{SS}^2 \sin\theta d\theta d\phi.
    \label{eqn:ss_method}
\end{align}

Both PFSS and MHD models produce similar \citep{Riley2006} coronal fields which conform to the standard paradigm that open field lines are rooted in coronal holes, which are dark regions observed on the solar disk at extreme ultraviolet (EUV) wavelengths. This correspondence supplies an observational constraint on such models, namely, the assumption that the foot-points of open field lines must correspond to the observed EUV coronal holes. A good agreement, at least for PFSS models, usually requires the source surface height to be in the range of $1.8-2.5 R_\odot$ \citep[e.g.  ][]{Lee2011,Badman2020,Reville2020}.  A complementary method motivated by this required correspondence is to empirically measure coronal hole boundaries in EUV imagery and simply sum the photospheric flux within these contours; \citet{Wallace2019} showed such estimates are in good agreement with the modelled values of $\Phi_{open}$.

The second context in which open flux is estimated is with the use of collections of single point in situ measurements of the HMF. While a set of single-point measurements confined along a spacecraft trajectory may at first appear to constitute a very weak statement on the state of the whole heliosphere,  it is, in fact, extremely powerful due to two key symmetries: the first is that in the \citet{Parker1958} model of the HMF, the radial component of the magnetic field along a streamline ($\theta,\phi$) varies as $B_R(R,\theta,\phi) = B_0(\theta,\phi) (R_0/R)^2$, and, thus, the quantity $B_R R^2$ is independent of radius. The second is one of the seminal results \citep{Smith1995,Smith2003} of the Ulysses mission \citep{Marsden1986}, namely, that $B_R R^2$ is independent of latitude ($\phi$). Thus, the total unsigned magnetic flux threading a sphere at an arbitrary radius, $R_S,$ in the heliosphere, $\Phi_H$, may be computed as 

\begin{align}
    \Phi_{H}(R_S) = 2R^2 \int_0^{2\pi} |B_R(\phi,R=R_S)| d\phi,
    \label{eqn:hmf_method}
\end{align}

where the modulus sign is not strictly a correct operation due to the curvature of the Parker spiral. The modulus here really represents assigning the opposite sign to the flux contribution of field-lines in sunward (S) versus anti-sunward (AS) sectors. The longitudinal variation and this modulus operation are typically approximated with an (as yet ill-defined) `averaging' procedure, ultimately giving:

\begin{align}
    \Phi_{H} = 4\pi R^2 <B_R(R=R_S)>, 
    \label{eqn:hmf_method_lon_int}
\end{align}

which illustrates that single-point measurements of the HMF (in particular the radial field) are actually extremely powerful and in principle allow an estimate of the magnetic flux in the heliosphere. We note that we have purposefully introduced a new symbol to refer to the flux in the heliosphere ($\Phi_{H}$) in order to allow for the possibility that deviations from the Parker spiral model may result in $\Phi_{open} \neq \Phi_{H}$. We will also, throughout this paper, normalise these $\Phi$ quantities (with nominal units of Webers) by $4\pi$ (1AU)$^2$, which transforms them to units of nT AU$^2$ and gives them the intuitive meaning of `magnetic flux density at 1AU'; the magnitude of which is almost exclusively confined to the range $0-10$ nT AU$^2$.

The possibility that  $\Phi_{open} \neq \Phi_{H}$ is motivated by observations by \citet{Owens2008} of a possible enhancement of $\Phi_{H}$ with radius, related in past studies to the effect of velocity shears causing warps in the magnetic field \citep{Lockwood2009a,Lockwood2009b,Lockwood2009c}, and more general local field inversions identified by intervals of sunward electron heat flux \citep{Owens2017}, which have been shown to become more prevalent with the distance from the sun \citep{Macneil2020}. Such an overestimation is a compelling possibility due to the open flux problem, which could, in principle, be explained if the overestimation is large enough. 

On the other hand, a similar enhancement with radius has been shown to occur artificially due to issues with the definition of the `averaging' procedure \citep{Smith2011}. Typically, averaging for the purposes of computing $\Phi_H$ has two critical components. The first is a `pre-averaging' timescale which is performed on the raw, signed $B_R$ data to produce a base data product. The longer this timescale, the more magnetic sectors cancel out and the distribution of $B_R$ gets closer to zero. To some extent this is a physical operation to get rid of very rapid fluctuations but at some ill-defined timescale will result in artificially low values of $B_R$ - it is clearly desirable for any open flux estimation method to be independent of averaging timescale. The second step \citep[e.g.  ][]{Owens2008} is to arrest further cancellation by, for example, taking the modulus and averaging this over Carrington rotations to obtain a longitudinal average. This modulus operation, in particular, due to the rectification of inverted field when taking the modulus of $B_R$ was shown by \citet{Smith2011} to result in an increase in apparent values of $B_R R^2$ with increasing radius, even in the fast polar coronal hole wind measured by Ulysses at solar minimum, under which conditions kinematic effects and local field inversions are less prevalent.

\citet{Erdos2012} and \citet{Erdos2014} proposed avoiding the rectification issue by utilising 2D vector distributions of the field and identifying that these populations are bi-modal and corresponding to anti-sunward and sunward sectors aligned along the Parker spiral. By measuring the field strength along this Parker spiral direction before projecting into the radial direction, these authors showed that the excess flux measured at Ulysses as compared to 1AU data over two solar cycles was much reduced and the corrected values followed the same large-scale variation with solar cycle as measured at 1AU.  Throughout this paper, we refer to this technique as the `Parker spiral method' (PSM). It is important to note, however, these authors still made the choice to use a `pre-averaging' timescale of six hours for their time series of $B_R$ (and orthogonal Cartesian component $B_T$) prior to making this Parker spiral projection. 

\citet{Owens2017} investigated the impact of pre-averaging on this method and standard averaging of $|B_R|$, in addition to comparing them both to the kinematic and local inversion methods on 1AU data. They showed that this `pre-averaging' timescale affected the PSM-derived values of $\Phi_H$ as well as for standard averaging. \citet{Owens2017} also showed, with the caveat that all their data was from 1AU, that:\ (1) the local inversion method was equivalent to the $|B_R|$ and PSM methods performed on pre-averaged data at a 1 day timescale; (2) with smaller pre-averaging timescales, the PSM method produced lower estimates for $\Phi_H$ than the $|B_R|$ method, while for higher pre-averaging timescales, the two methods produced similar estimates. The reason for the dependence of the PSM value on  the pre-averaging timescale is due to a very important subtlety with regard to all the implementation of the PSM and pre-averaging on the part of the authors above. Namely, the pre-averaging they employ is applied to the 3D HMF vector in Cartesian coordinates and the resulting averaged-down Cartesian time series are what is used to perform the projection onto the Parker spiral. However, as explored in Appendix \ref{sec:pre-av}, averaging in the Cartesian basis strongly distorts the 2D distribution of vector field measurements due to fact that the Cartesian components are correlated by fluctuations, which are predominantly spherical in nature. This issue can be better addressed by parameterising the vector field in polar coordinates before any pre-averaging is done. We return to this point when we introduce our implementation of the PSM in Section \ref{sec:2d_vs_r}.

Whether artificial or physical, establishing whether the measurements of $\Phi_H$ is an overestimation is of key importance for the open-flux problem in order to detect whether $\Phi_H = \Phi_{open}$ and, thus, whether this can serve as a real constraint on coronal models. In this paper, we aim to utilise new Parker Solar Probe (PSP) observations at an unprecedented range of heliocentric distances, together with corresponding observations at 1AU, to study the variation of $\Phi_H$ throughout the inner heliosphere, assess the dependence on the processing technique, and determine how these measurements contribute to the discussion of the open flux problem. In Section \ref{sec:obs}, we introduce the data set and the different methods that are used to estimate $\Phi_H$. We utilise synthetic data to examine how these different methods are affected by radial trends in the background field and fluctuations. In Section \ref{sec:results} we show the results of measurements of $\Phi_H$ throughout the inner heliosphere, assess the possible contribution of inverted flux to these measurements and compare them with estimates of the open flux derived with PFSS models. In Section \ref{sec:disc}, we discuss the implications of these results. Finally, in Section \ref{sec:concl} we summarise our main conclusions.

\section{Observations and methods}
\label{sec:obs}

\subsection{Spacecraft data summary}
\label{sec:data}

\begin{figure*}[h!]
    \centering
    \includegraphics[width=\textwidth]{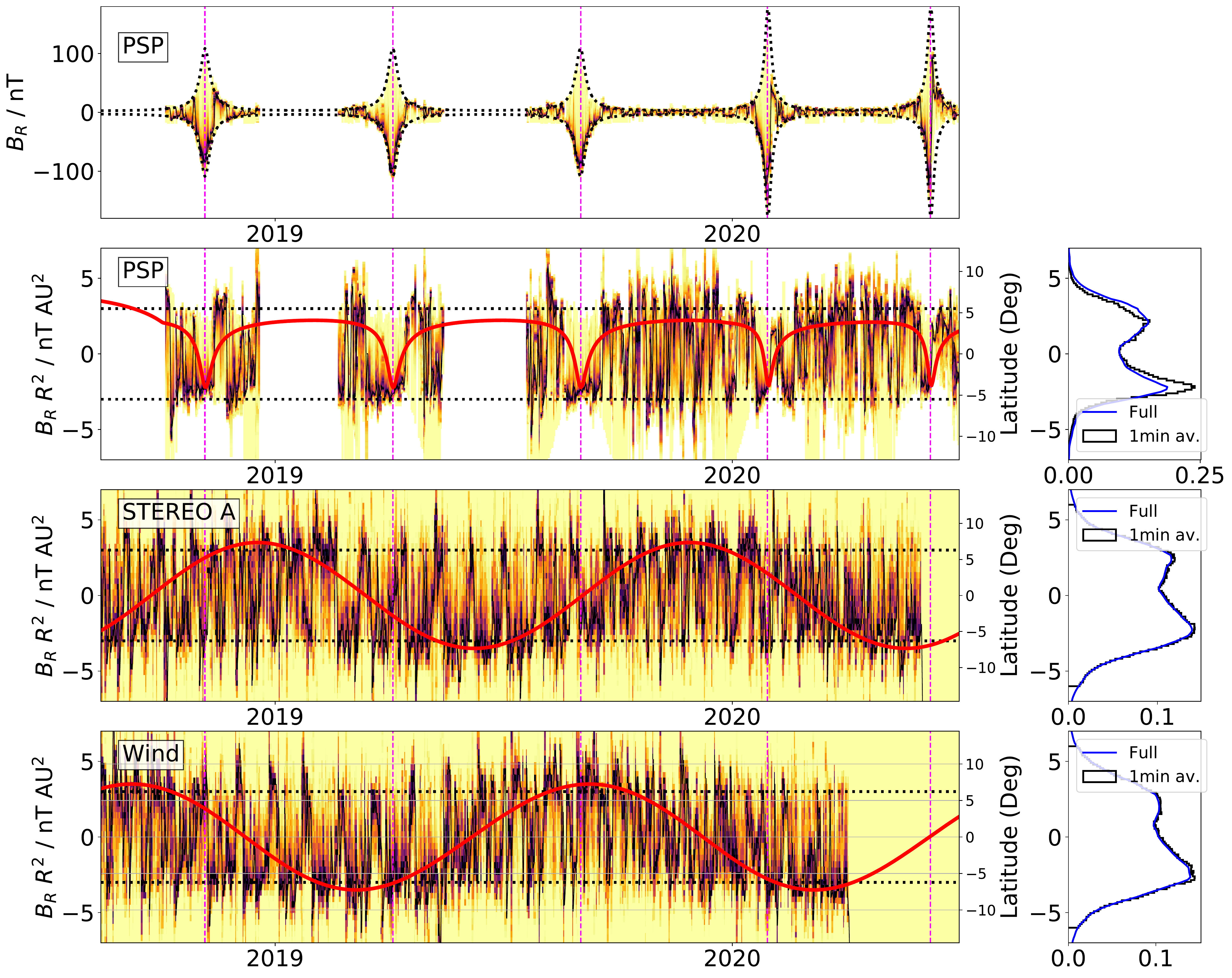}
    \caption{Summary of the in situ magnetic field data analysed in this paper. In each panel, the shading is a 2D histogram with the x-axis in 1 day bins. A black solid line threads the histograms and shows the mode of each day. The top panel shows the raw radial magnetic field measured by PSP. Magenta dashed vertical lines indicate successive perihelia of PSP. A faint dotted black line indicates an envelope (3nT (1AU/$R^2$) ) which communicates PSP's changing heliocentric distance. The remaining three rows show the quantity $B_R R^2$ as measured by PSP, STEREO A and Wind respectively. The dotted horizontal line is the same envelope from the top panel scaled by $R^2$. In each panel, a solid red curve shows the spacecraft latitude. The polarity sampled is controlled primarily by the latitude. Panels on the right are 1D histograms which show in black (blue) the distributions of the 1 min average (full cadence) data of $B_R R^2$ summed over time.}
    \label{fig:synoptic}
\end{figure*}

 The Parker Solar Probe \citep[PSP; ][]{Fox2016} was launched on August 2018 into a highly elliptical heliocentric orbit with an inclination of $\sim 4^o$ to the solar equatorial plane. With a sequence of Venus gravity assists, PSP's perihelion distance decreases over the course of the mission. As of November 2020, PSP has completed six orbits with its first three perihelia at 35.7$R_\odot$ (0.166 AU), and its fourth and fifth both at 27.8$R_\odot$ (0.129 AU). The most recent perihelion was even closer at 20.4 $R_\odot$ (0.09 AU). Thus PSP provides measurements of the heliospheric magnetic field more than twice as close to the Sun as the previous record holder, Helios, at 65$R_\odot$ (0.3AU). In addition to this unprecedented radial evolution, PSP also samples longitudinal structure in a unique way: as the PSP approaches perihelion, its orbital velocity increases faster than the co-rotation velocity of the Sun and, as a result, it crosses a threshold, where it first co-rotates and then super-rotates with respect to the solar photosphere. The upshot of this is that PSP samples longitudinal variation very slowly, with two intervals of co-rotation (inbound and outbound) where it measures the same solar wind for an extended period. The downside to this is that each PSP perihelion probes only a small range of solar longitude, although data from its cruise phase at larger heliocentric distances provides measurements all around the Sun.
 
 In this work, we utilise DC magnetic field data from the FIELDS instrument \citep{Bale2016} \footnote{Data publicly available at \url{https://fields.ssl.berkeley.edu/data/}} from orbits 1-5 of PSP, spanning from August 2018 to July 2020 (data from orbit 6 will soon be available and allow for a further extension of the present study). We utilise the full time-series of the one minute (1 min) averaged data product (B\_RTN\_1min), in addition to producing histograms of the four-samples-per-cycle ($\sim$4.6 Hz) data product (B\_RTN\_4\_per\_cyc) over hour-long and day-long timescales. These data utilise the inertial Radial-Tangential-Normal (RTN) coordinate system, with the R component indicating the radial direction at the spacecraft position, the T component defined as the cross product of the radial direction with the solar rotation axis, and the N component completing the orthogonal triad. In terms of heliographic coordinates, T points along a line of constant solar latitude ($\phi$) and N points along lines of constant solar longitude ($\theta$). While $B_R$ is the main vector component of interest in this paper, as discussed at length in Sections \ref{sec:2d_vs_r}-\ref{sec:measured_trends}, knowledge of the full 3D vector is key to understanding the evolution of individual components. Here, we also introduce the angular quantity $\alpha,$ which we refer to as the `clock angle', which is the angle the HMF vector makes with the radial direction when projected into the R-T plane.
 
 The data from PSP and the near-1AU spacecraft considered in this study are summarised in Figure \ref{fig:synoptic}. Here, each panel is a 2D histogram of data value on the y-axis and time on the x-axis. The time axis is binned into one day intervals and each column of histogram data shows the distribution of the full resolution data set for each of these intervals.  A solid black line threads the modal value of the data and indicates which side of the heliospheric current sheet (HCS) PSP was located on each day. Dashed magenta lines indicate the times of perihelia and allow reference to the other panels. The colour scale of the histograms (light to dark) shows the density of measurement (low to high). The top panel shows PSP's measurements of the radial component  ($B_R$) of the Heliospheric Magnetic Field (HMF). A dotted envelope showing $B_R = \pm 3 nT (1AU/R^2)$ bounds the data and reveals that, as expected, the predominant variation in this component is proportional to 1/$R^2$. It also highlights the lower perihelia distances of orbits 4 and 5, where the envelope and data range gets larger than during orbits 1 through 3.  We can also clearly see that no matter which side of the HCS the PSP is located on any given day, there is a non-negligible population of field measurements of the opposite polarity. Particularly at the closest approach, this corresponds to the prominent switchbacks, which were a key early discovery of the PSP \citep[e.g.  ][]{Bale2019, Kasper2019, DudokdeWit2020, Horbury2020}. The remaining rows show the quantity of principal interest in this paper, $B_R R^2$, displayed in the same histogram format over the investigated time interval as measured by PSP/FIELDS, the IMPACT/MAG \citep{Luhmann2004} instrument on STEREO AHEAD \citep{Kaiser2004} and the MFI investigation \citep{Lepping1995} on the \textit{Wind} \citep{Harten1995} spacecraft. In these latter three panels, a solid red line shows the variation of each spacecraft's heliographic latitude with time. We note PSP's minimum latitude and minimum heliocentric distance are closely related. 
 
 Finally, to demonstrate the relation between the full cadence and one minute (1 min) average data products used in this work, 1D histograms of both data products are shown overlaid in a series of inset panels on the right. For 1AU, the distributions are virtually identical. For PSP, the one minute average distribution is slightly distorted compared to the four-samples-per-cycle data. In particular, the positive tail of values is foreshortened by the 1 min averaging data but the peak values are unchanged. Thus, we look to use the more pristine faster cadence data product where possible, however, we argue the distributions are similar, especially with regard to the data set of negative polarity, such that using the 1 min average data where necessary will not present an issue. For computational tractability, the 1 min averaged data will be used when considering the whole set of five orbits, while the higher cadence data will be used when considering one-hour or one-day intervals of data.
 
 From the top panel, the data coverage is immediately apparent, demonstrating that there is no PSP/FIELDS data outside of the `encounter' phases of encounters one, two, and three, while for orbits 3 to 5, there is continuous data coverage. At all three spacecraft, there is a striking correlation between latitude variation and dominant measured polarity: When the spacecraft are at their minimum (maximum) latitudes, they are more likely to be southwards (northwards) of the HCS and, therefore, more often measuring negative (positive) polarities for $B_R$. For STEREO A and Wind, whose orbits have a constant angular velocity with respect to the solar co-rotating frame, the variation from two timescales is immediately apparent: slow annual latitude variation and regular HCS crossings due to the $\sim 27$ day Carrington rotation period. For PSP, at its closest approach, it can be seen that a single dominant polarity is measured for extended periods of time while PSP is close to co-rotation with the Sun, whereas for the aphelion periods between the latter orbits, the familiar Carrington rotation pattern reappears.
 
 A dotted line at $\pm$ 3nT AU$^2$ in the latter three panels shows the same envelope as in the top panel with the radial variation scaled out. Zooming in this way, we see the 1AU spacecraft regularly measuring values of $B_R R^2$ exceeding this value, while PSP's measurements are generally below this value, especially during the perihelia. While it is clear $B_R R^2$ is approximately of constant magnitude and similar in value at all three spacecraft, there is significant scatter and temporal variation. The remainder of this paper is concerned with analysing these distributions of $B_R R^2$ and the full HMF vector to justify a statistical `background' value and obtain a best estimate of $\Phi_H$. We assess the extent to which it is conserved throughout the inner heliosphere and, finally, how it compares to estimates from PFSS models. 

\subsection{Radial evolution}
\label{sec:radial_evolution}   

\begin{figure*}[h!]
    \centering
    \includegraphics[width=0.9\textwidth]{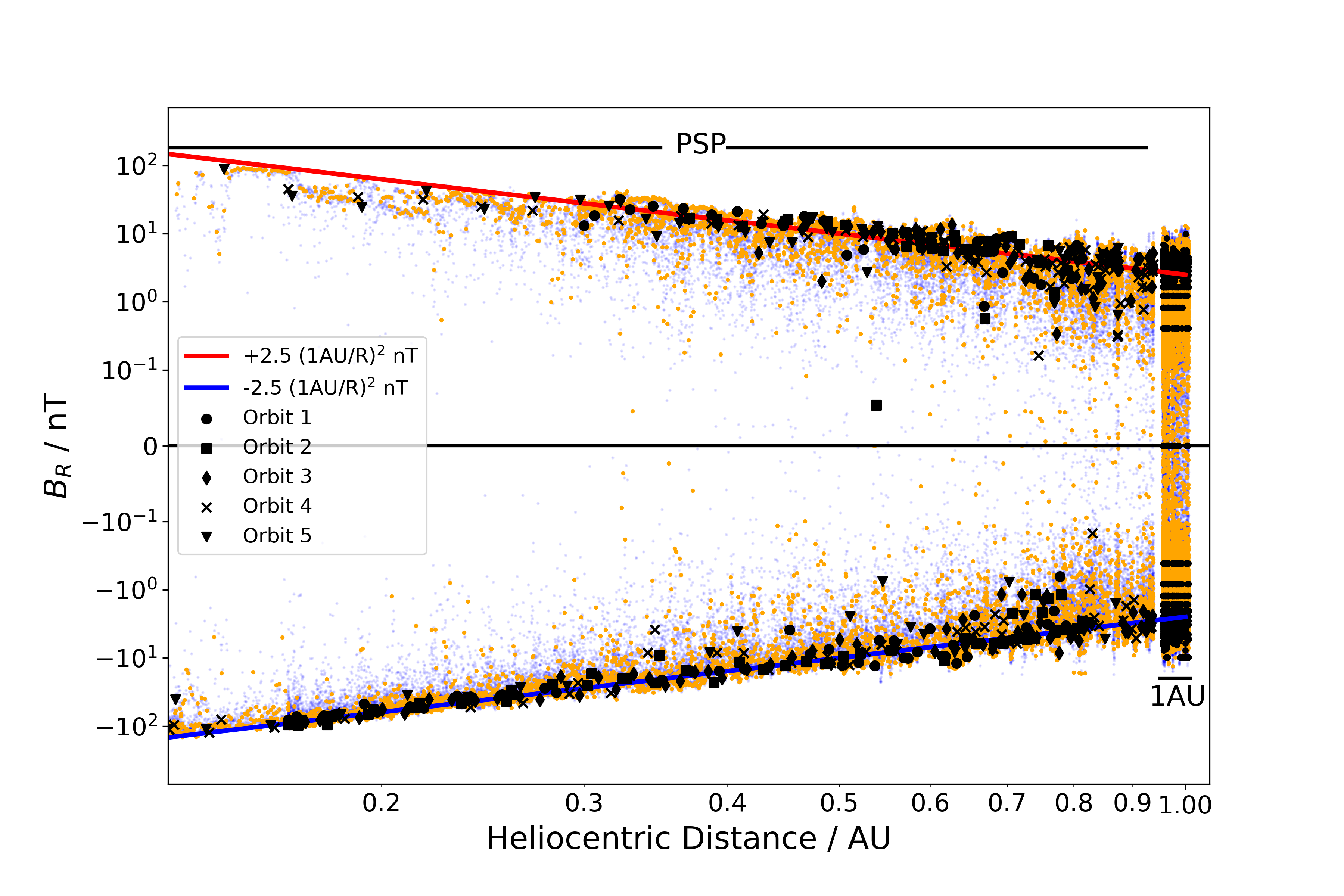}
    \caption{Radial scaling of $B_R$. All the data from Figure \ref{fig:synoptic} is re-plotted against radius on a symmetric log - log scale. Data with $|B_R| < 0.1 nT$ is plotted on a linear scale, which accounts for the block of data passing through zero near 1AU. One-minute averages are plotted in faint magenta. One-hour modes are shown in orange. One-day modes are shown in black with different symbols differentiating the subsequent orbits. We note that orbits 4 and 5 (crosses and triangles) extend to lower radii than orbits 1 to 3. A 2.5nT (1AU/R)$^2$ trend line is plotted for positive (red) and negative (blue) polarities. A small data gap shows the narrow gap in radial coverage between PSP’s aphelia and STEREO A’s perihelia.}
    \label{fig:br_vs_radius}
\end{figure*}

We first consider the radial variation of $B_R$ directly to confirm the $1/R^2$ transformation is appropriate for the full data set. The measurements as a function of radius are plotted in Figure \ref{fig:br_vs_radius} on a log-log scale. Faint blue dots show the spread of the 1 min averages of $B_R$ from PSP's first five orbits (the lower cadence `raw' data), while orange dots show the most probable value of $B_R$ for each hour, demonstrating the most probable values of the raw data cluster on a 1/R$^2$ trend line. The black markers indicate the most probable values for each day and the different shaped markers represent the different orbits and are shown to include a data product which is sparse enough for the reader to see the difference between the different PSP encounters. Solid red and blue lines show $\pm 2.5 nT (1AU/R)^2$ trend lines. While these trend lines are not fitted and serve here only as a visual aid, as we show in Section \ref{sec:results}, $\Phi_H = B_R R^2 = 2.5\text{nT AU}^2$ actually turns out to be the best estimate for this data set. It is worth noting that while the 1AU data appears to be significantly more spread than the PSP data, this is mostly a distortion due to the combined effects of the higher data density resulting from the limited radial variation of these spacecraft, the symmetric log scale, and the population of vectors where the full field vector fluctuates into the radial direction, such that $B_R = |B|$ instantaneously; at 1AU, $|B|$ is significantly larger than the background $B_R$ value.

This effect aside, we observe: (1) the radial component of the field indeed varies as 1/$R^2$ for all radii probed by PSP to date. (2) the scatter around the trend line is proportionally smallest at the closest in heliocentric distances. It is also systematically skewed towards zero but this is not trivial to observe on the log scale. (3) Orbits four and five show a significant departure from the trend line towards weaker field strengths, particularly on the positive field branch around 0.2AU. This is likely due to PSP running very close to a very flat heliospheric current sheet during these latter orbits (Chen et al., submitted to this issue), and thus sampling more weakly magnetised streamer belt plasma. We conclude using (1), that the quantity $B_R R^2$ is a useful quantity which is, at least approximately, conserved throughout PSP's mission so far. There is no suggestion of a break down of the Parker expectation or the Ulysses result of latitudinally isotropised radial magnetic field and suggests the dipole-dominated coronal field has fully relaxed into this isotropic state, which is well within PSP's closest heliocentric distance of $28 R_\odot$.

\subsection{Radial evolution of the distribution of magnetic flux}
\label{sec:2d_vs_r}

\begin{figure*}[h!]
    \centering
    \includegraphics[width=\textwidth]{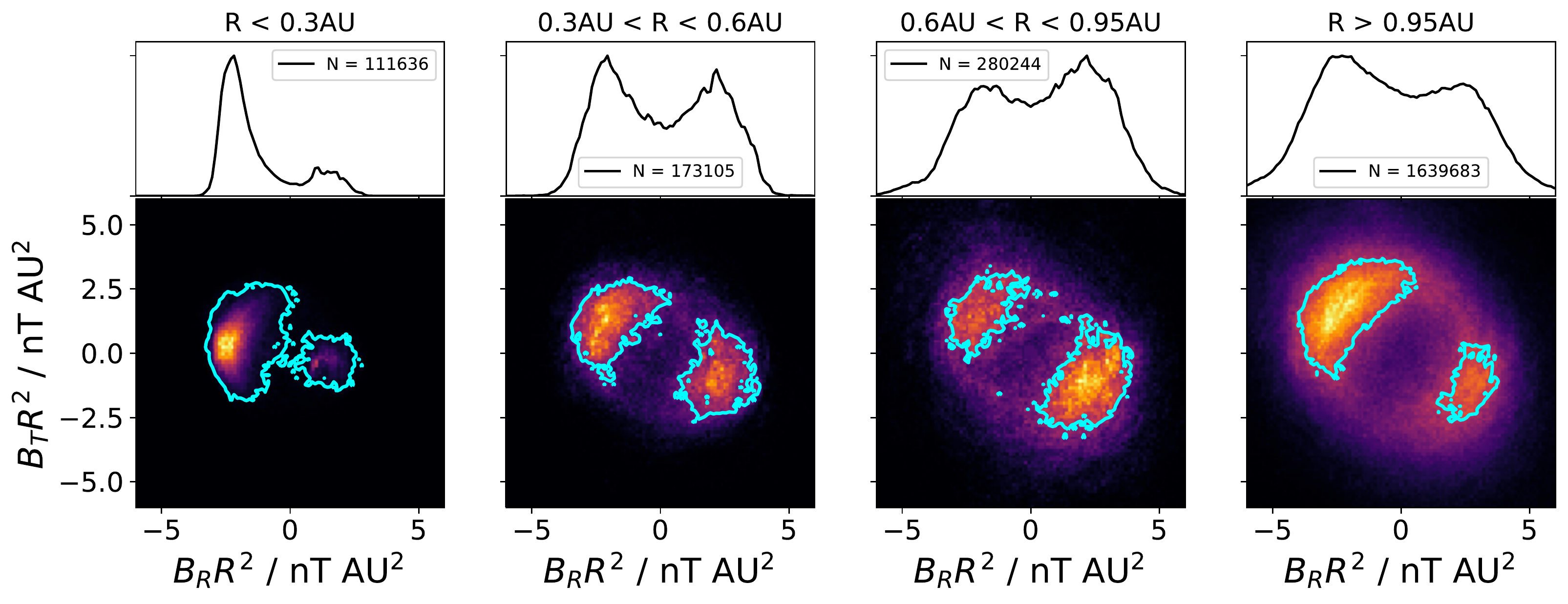}
    \caption{Radially evolving distribution of $B_R R^2$.  Each panel in the bottom row contains 2D histograms of $B_T R^2$ versus $B_R R^2$, which show the distribution of the field in the RT plane in a given radial bin, indicated for each column. The cyan contour depicts the 90th percentiles of the data. The top row show the resulting 1D distribution of $B_R R^2$. The legend gives the number of data points (1 min averages) in each radial bin. We note the left three columns are from PSP data, while the right hand panel is the summation of two years of Wind and STEREO A data}
    \label{fig:2d_dists}
\end{figure*}

With the investigation of the quantity $B_R R^2$ remaining well-motivated, we turn our attention to measuring it directly. A key assumption made in displaying the data above, as we have done so far, is to show the most probable value of the field. That is, we take all the data from some time interval or other binning procedure, produce a distribution of that binned data, and assume the peak of that distribution (the bin with highest number of counts) represents the value we are trying to measure. 

In Figure \ref{fig:2d_dists}, we demonstrate how the field is distributed in the R-T plane at different heliocentric distances. We very coarsely bin all the data shown in Figure \ref{fig:synoptic} into four categories: PSP data for R $<$ 0.3 AU (specifically, PSP data interior to the closest perihelia of Helios), PSP data for 0.3AU $<$ R $<$ 0.6AU, 0.6 $<$ R $<$ 0.95 AU, and finally, all the STEREO A and Wind data (for which R > 0.95 AU). The data shown here reflect the 1 min averages due to the aforementioned computational tractability issues when working with the whole data set. The top panel for each radial bin shows the 1D distribution of $B_R R^2$, the bottom panel shows 2D distributions of $B_T R^2$,$B_R R^2$. We note that we have made a replacement here, namely, $B_R = |B| \cos \alpha$ and $B_T = |B| \sin \alpha$. Here, $|B| = \sqrt{B_R^2 + B_T^2 + B_N^2}$ is the field magnitude and $\alpha = \arctan2 (B_T,B_R)$ is the field vector angle of rotation in the R-T plane relative to the radial direction, and will be referred to as the `clock angle'. This transformation is discussed in Appendix \ref{sec:appendix:normal_flucs}. It is done to avoid the effect of projecting fluctuations in the normal direction onto the R-T plane, which can lead to an underestimation in field magnitude. In particular, here we are assuming the normal, tangential, and compressive fluctuations are uncorrelated, and so making this correction to the R-T components does not affect the $|B|$ and $\alpha$ distributions.  

We also note that in Figure \ref{fig:2d_dists}, the two peaks values are asymmetric. This is simply a sampling effect based on the spacecraft orbit (especially the heliographic latitude). For example, PSP's orbit is tilted with respect to the solar equatorial plane such that at closest approach, it is also approximately at its minimum (and most negative) latitude. For this reason, PSP for R < 0.3 AU is primarily southwards of the HCS and, therefore, it samples a negative-polarity magnetic field. We can confirm this via an inspection of the second panel of Figure \ref{fig:synoptic} in which the dips where PSP goes into negative latitudes correspond to protracted measurements of negative polarity.

Figure \ref{fig:2d_dists} shows that the 1D distributions of $B_R R^2$ are the projection of a 2D bi-modal distribution which evolves with radius. This 2D distribution is aligned with a mean Parker spiral direction which becomes closer to radial with decreasing heliocentric distance. The distributions comprise of a sunward (upper left quadrant) and anti-sunward (lower right quadrant) population, and exhibiting some spread in field magnitude ($|B|$) and angle ($\alpha$) about this mean state. Empirically, these spreads decrease with heliocentric distance and this trend will be quantified in Section \ref{sec:measured_flucs}. Further from the Sun, an overlap of the two populations sum together to distort the 1D distribution of $B_R R^2$ by producing a large population of values with $B_R R^2$ close to zero. This effect is much weaker close to the Sun, with the two peaks in the 1D distribution much sharper and better isolated. 

In this work, we use distributions of raw (meaning minimally pre-averaged) $B_R R^2$ and $B_T R^2$ measurements, such as those illustrated in Figure \ref{fig:2d_dists}, to produce estimates of the heliospheric flux. We consider three particular methods: the mean, the mode, and the Parker spiral method. The mean and mode are performed directly on the 1D $B_R R^2$ distributions. The mode simply identifies the bin with the most counts for either $B_R > 0$ or $B_R < 0$. For the mean, we take the bi-modal 1D distributions and bifurcate it into two truncated distributions, $B_R > 0$ and $B_R < 0$, and the mean of each is taken. At certain points in this work, we  consider one-hour or one-day intervals of data. For these cases, the distributions are usually single-peaked (only one magnetic sector explored) and, therefore, the procedure we follow is to identify the single most probable value for that hour or day and then we use the sign of this value to decide which side of the distribution to truncate. The Parker spiral method utilises the full 2D distribution, which we introduce and describe below.

Considering the 1D distributions we see at all radii, the shape of each half of the distribution is asymmetric (or `skewed'), meaning that the production of a best estimate of the quantity is not obvious given the mean and mode are non-trivially related (and the dependence of their relationship on radius). The mean in particular is sensitive to the `overlap' of the two sectors. In the simplest approach \citep[e.g.  ][]{Owens2008, Linker2017, Wallace2019}, the modulus of the distribution is taken before computing the mean, which is similar to the truncation procedure we described above. In this method, the larger the `overlap' population relative to the peak, the higher the mean. Given that this overlap population grows with heliocentric distance (see Figure \ref{fig:2d_dists} above and also \citet{Erdos2012}), this may cause the apparent heliospheric flux to grow with radius when this method is used \citep{Smith2011}.  

In a more careful treatment of the bi-modal population, we would need to subtract the population of the opposing sector, for example, by fitting a curve to the distribution, or to make another approximation, such as bisecting the data in the 2D distribution with an approximate Parker spiral, and to assume that no data fluctuates past 90 degrees from the mean Parker spiral direction \citep{Erdos2014}. An even more sophisticated approach would be to use the electron heat flux to delineate these two populations such that intervals where the electron heat flux is parallel (anti-parallel) to the magnetic field correspond to anti-sunward (sunward) sectors (see Section 3.2 of \citet{Macneil2020}). As mentioned at several points in this work, this application of the electron strahl remains an important avenue for exploring additions to the analysis we describe here. We omit the usage of electron strahl measurements due to the more intermittent temporal coverage of solar wind plasma measurements by the PSP Solar Wind Electrons Alpha and Protons (SWEAP) instrument \citep{Kasper2016} as compared to the full FIELDS data set analysed here.

The mode is more robust in that it is less affected by the two sectors overlapping and has the advantage that it can easily be defined for both sunward and anti-sunward sectors. It is still possible the overlapping populations could change the peak value somewhat. The mode also requires a large sample size in order to be well-defined and can have a large error when computed from a distribution with a flattened peak. In all these regards, we see the mode is better defined for the distribution of $B_R R^2$ closer to the Sun, where the peaks are narrower and the impact of overlap from the positive sector is very weak. 

In light of these subtleties, \citet{Erdos2012} proposed utilising the apparent symmetry in the 2D distributions directly by projecting the data along the nominal Parker spiral direction, computing the mean of data along this direction and projecting this into the radial direction to obtain a best estimate of $\Phi_H$. As shown by the shape of the 2D distributions in Fig 3., this approach is supported by PSP data and even more so close to the sun where the peak in the 2D distributions become very sharply defined (even though, for this data set, they are predominantly in the sunward sector). Moving forward, we refer to this technique as the `Parker spiral method', or `PSM'. In this work, we use an empirically measured Parker spiral angle rather than the ideal angle generated with a solar wind velocity as done by \citet{Erdos2012}. This choice is made because, as mentioned above, the solar wind velocity data set is a subset of the magnetic field data set from PSP that is due to differing instrument operation schedules and constraints. Thus, it is desirable to use a method which can be performed purely with magnetic field observations. Furthermore, the Parker spiral is an idealised model and, in fact, driving a model with a varying solar wind velocity leads to some unphysical inferences, such as slower wind streams overlapping with faster ones (which form stream interaction regions in more realistic simulations). Additionally, recent observations by PSP \citep{Kasper2019} call into question the Parker spiral assumption that the solar wind velocity flow is purely radial at distances probed by PSP so far. By empirically estimating the background vector field, we avoid the need to assume an exact Parker model or to use more sophisticated models. Finally, an empirical Parker spiral angle allows us to work directly with the vector in a spherical representation when implementing the PSM, which is preferable to the Cartesian representation, as described below.

To implement the PSM, we form 1D distributions of |B| and the clock angle ($\alpha$) which underlie the 2D distributions in the R-T plane shown in Figure \ref{fig:2d_dists} and measure the most probable value of these quantities, $B_0$, $\alpha_0$. When the distributions are bi-modal, a most probable sunward and anti-sunward value of $\alpha_0$ is found. The most probable values are computed by identifying the bins of these 1D distributions of $|B|$ and $\alpha$ with the most counts. Then the estimate of $\Phi_H$ for the PSM is derived as $\Phi_H = B_0\cos \alpha_0 R^2$ where $R$ is the spacecraft heliocentric distance relevant to that distribution. The separation of these 2D distributions into 1D magnitude and angle distributions are illustrated in Figures \ref{fig:appendix:synth_vs_real_close} and \ref{fig:appendix:synth_vs_real_far} in Appendix B.

Is is important to discuss the effect of pre-averaging on our different estimation methods. When the HMF vector field is pre-averaged in Cartesian coordinates, the whole 3D distribution is deformed by more and more aggressive time-averaging and this procedure affects all the methods, including the PSM and the mode. In particular, the effect on bi-modal distributions is to shift the peaks towards zero, causing estimates of the heliospheric flux to decrease, as observed for all methods by \citet{Owens2017}. In terms of the 2D distributions depicted in Figure \ref{fig:2d_dists}, the effect of longer pre-averaging on the raw data would be to shrink the whole annulus of data towards the centre (see Appendix \ref{sec:pre-av}). In terms of quantifying this, \citet{Owens2017} showed one hour and one day averages produced significantly lower values of $\Phi_H$ (a reduction of $\sim 20 \%$, although this only applies to 1AU data) and we should note that \citet{Erdos2012} used a six-hour average for their PSM implementation. In this work, we attempt to avoid pre-averaging issues primarily by applying our estimation methods to distributions of 1 min averaged or higher cadence data such that our vector distributions are minimally distorted. However, we also note one very important characteristic of our PSM implementation is that we parameterise our vector field in spherical coordinates at very high time resolution. The resulting parameterised distributions are evenly distributed about a central value for most radii. This means that as long as the field stays in one sector (sunward or anti-sunward) for most of each averaging window, these time series of $|B|$ and $\alpha$ should be able to be arbitrarily "pre-averaged" without affecting the results. This is demonstrated with STEREO A data in Appendix \ref{sec:pre-av}. As mentioned above, use of the electron strahl would be a robust way to delineate the sectors a given HMF measurement belongs to.

We reiterate here that in certain parts of this work, our flux estimation methods are applied:  (1) to data over one-hour intervals; (2) data over one-day intervals; (3) data from the full mission binned by heliographic location. For case (3), there are sufficient data points using 1 min averages to establish robust histograms to study. For cases (1) and (2), we utilise the four-samples-per-cycle PSP/FIELDS data product, providing N = 16500 data points for each hour and over 400,000 for each day, ensuring well-formed histograms for these cases too. We next move on to compare and contrast our different estimation methods introduced above, starting by applying them to an idealised synthetic data set.

\subsection{The impact of vector fluctuations on the flux distribution}
\label{sec:synthetic}

\begin{figure*}[h!]
    \centering
    \includegraphics[width=\textwidth]{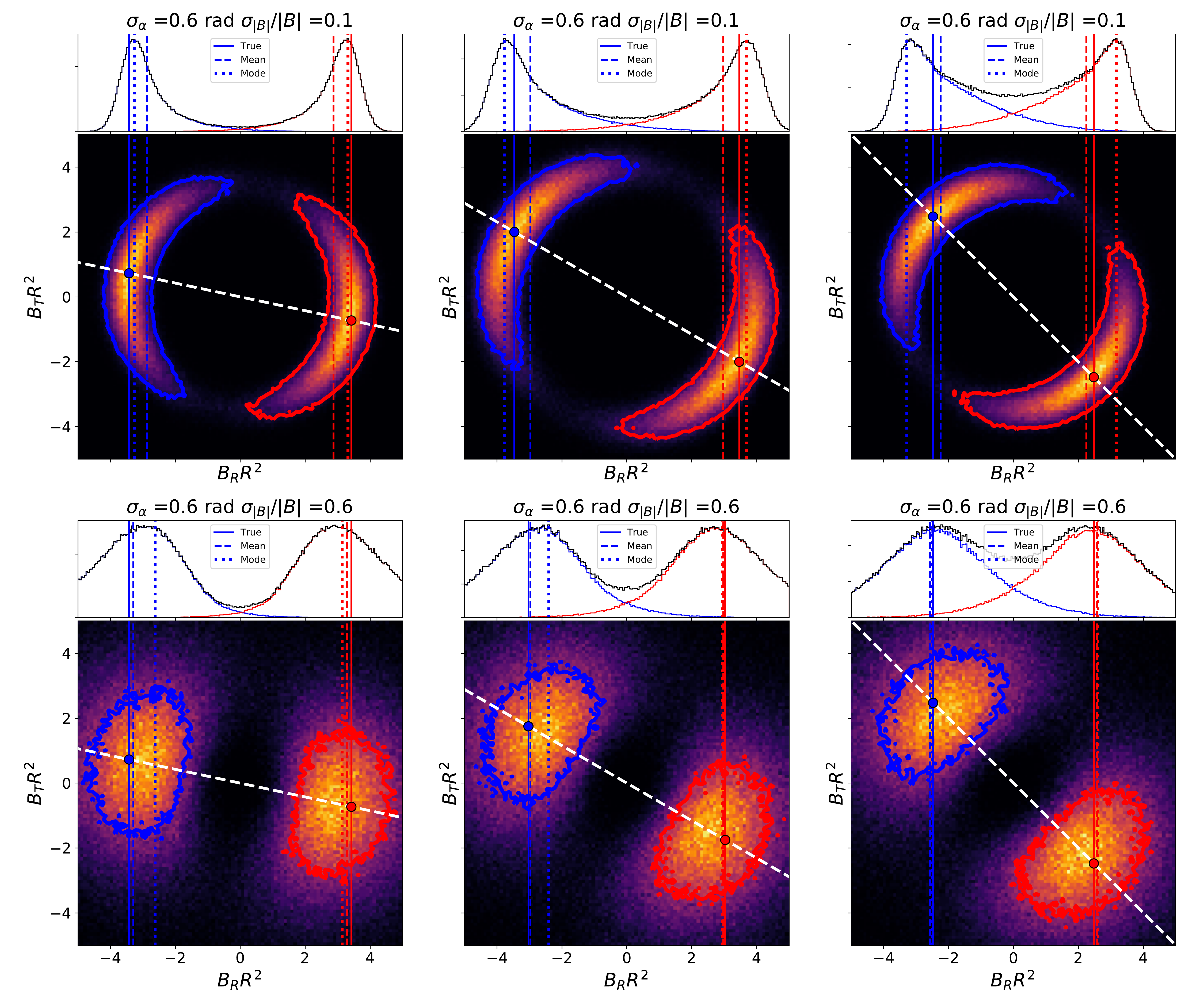}
    \caption{Synthetically constructed 2D and 1D distributions of HMF vector measurements and resulting statistics. Similarly to Figure 3, each panel shows a 2D synthetic distribution of $B_T R^2$ - $B_R R^2$ drawn (see main text) from a mean value and standard deviation in clock angle ($\alpha$) and field magnitude ($|B|$). A separate distribution for sunward (S) and Anti-sunward (AS) sectors are drawn. The colour-map shows the full distribution, and blue and red contours show the 90th percentile of the S and AS sectors. A white dashed line depicts the mean clock angle. The text above each panel describes the standard deviation in angle in radians and field magnitude normalised by mean field (we note that both are dimensionless such that their relative balance is apparent). Red and blue circles indicate the AS and S central ('true') values, respectively. Above each panel is the resulting 1D distribution, with blue and red curves showing the individual distributions and black showing the joint distribution. Solid, dotted, and dashed lines show the results of the PSM, the distribution mode, and the distribution mean as measured from that distribution, respectively. The top row shows the case for fluctuations in the clock angle dominating over fluctuations in magnitude, while the bottom row shows balanced fluctuations. From left to right, the mean clock angle increases from 12$^o$ to 45$^o$, which is approximately the range of angles probed by PSP as its heliocentric distance varies.
}
    \label{fig:synthetic}
\end{figure*}

To investigate how the underlying 2D distribution of magnetic field data affects the statistics of the 1D distributions of $B_R R^2$ and how this affects our different ways of measuring $\Phi_H$ introduced in the previous section, we begin by producing a synthetic data set. Motivated by \citet{Erdos2012}, we suggest that the 2D field be modelled in a polar representation as a set of normally distributed fluctuations in vector magnitude ($|B|$) and R-T clock angle ($\alpha$) with standard deviations $\sigma_{|B|}$ and $\sigma_\alpha,$ respectively, which may be thought of as encoding the relative balance of compressive and rotational (or Alfv\'enic) fluctuations respectively. It is important to state we are making a strong assumption that fluctuations are evenly distributed about an average value which also corresponds to the mode and the `central value'. This assumption allows us to establish a set of quantities ($|B_0|$, $\alpha_0$, and $B_{R0} = |B_0|\cos \alpha_0$), which can be defined as a `ground truth', and which the PSM measures directly by our method of constructing this synthetic data.

In Figure \ref{fig:synthetic}, we generate ensembles of synthetic measurements by drawing from normal distributions using the \url{random} module in the \url{numpy} library \citep{harris2020array} in python over a range of background clock angles and fluctuations, and project them into synthetic 1D $B_R R^2$ distributions as in Figure \ref{fig:2d_dists}. Because we generate the sunward and anti-sunward sectors separately, we can keep track of the individual distributions (red and blue curves) and see how they add up to give the overall distribution. It is immediately apparent that the overlap grows with radius simply because of the increase in the Parker spiral angle \citep[c.f. ][]{Erdos2012,Erdos2014}.

We note that for a central value of $|B_0|$, the dimensionless quantity $\sigma_{|B|}/|B_0|$ is a useful figure of merit which may be compared to $\sigma_\alpha,$ as expressed in radians to describe the balance of fluctuations. In the top row of Figure \ref{fig:synthetic}, distributions dominated by rotational fluctuations at various background clock angles ($\alpha_0$, as indicated by the white dashed line) are shown, while in the bottom row, compressive and rotational fluctuations are balanced. By comparing Figures \ref{fig:2d_dists} and \ref{fig:synthetic}, we can see that the `boomerang' shapes of fluctuations dominated by rotations are heuristically more similar to the real PSP data. This will be established quantitatively in the next section.

As expected, the Parker spiral method exactly reproduces the `ground truth' as indicated by the blue and red circles aligned with the spiral direction in the 2D distributions. The solid (red or blue) vertical lines indicate the value of $\Phi_H$ that would be measured by the Parker spiral method, and that link the 2D and 1D distributions. Dotted and dashed vertical lines respectively respectively show the measured mode and mean based on the 1D summed distribution (black curve). The mean for both positive and negative sectors is computed by truncating the distributions at $B_R R^2 = 0,$ as discussed in Section \ref{sec:2d_vs_r}. To be clear, these operations are applied directly to the distributions depicted here which represent the `pre-averaged' or base data product. At all points in dealing with real data, the maximally pre-averaged data we use are the 1 min averages.

By comparing the vertical lines in Figure \ref{fig:synthetic}, we see that our three methods give slightly different results and the specific differences depend on both the nature of the fluctuations and the mean spiral angle. In particular, when rotational fluctuations dominate, the mean value is an underestimation, especially when the Parker spiral is close to radial (left-most column). At these spiral angles, the mode is also a slight underestimation; however, at a higher Parker spiral angle (right-most column), the mode appears to overestimate the true value. The skewed 1D distribution is being primarily shaped by the geometry of projecting angular fluctuations and so, the distribution peak does not in general correspond to the 2D background value about which the field vector is fluctuating. Thus, the assumption that the mode is `more representative' than the mean is not yet justified.

When the compressive and rotational fluctuations are balanced, all three values are closer together (the mean is a better approximation), especially at a higher Parker spiral angle. In this case, for low Parker spiral angles, the mean is, in fact, higher than the mode. 

Overall, in this section, we show that the nature of the 2D fluctuations of the vector and the inclination of the Parker spiral angle can affect our measured value of the heliospheric flux. Therefore, if the vector fluctuations can indeed be modelled as normally distributed about a mean magnitude and angle -- especially if rotations dominate over vector magnitude fluctuations and then the Parker spiral method is clearly the most robust and physically motivated method considered here. The fluctuation balance would also determine the appropriateness of each method. Next, we seek to measure the background vector and fluctuation statistics as a function of radius from the real spacecraft data and we also check the extent to which fluctuations are evenly distributed.

\subsection{Measured fluctuations}
\label{sec:measured_flucs}

\begin{figure*}[h!]
    \centering
    \includegraphics[width=\textwidth]{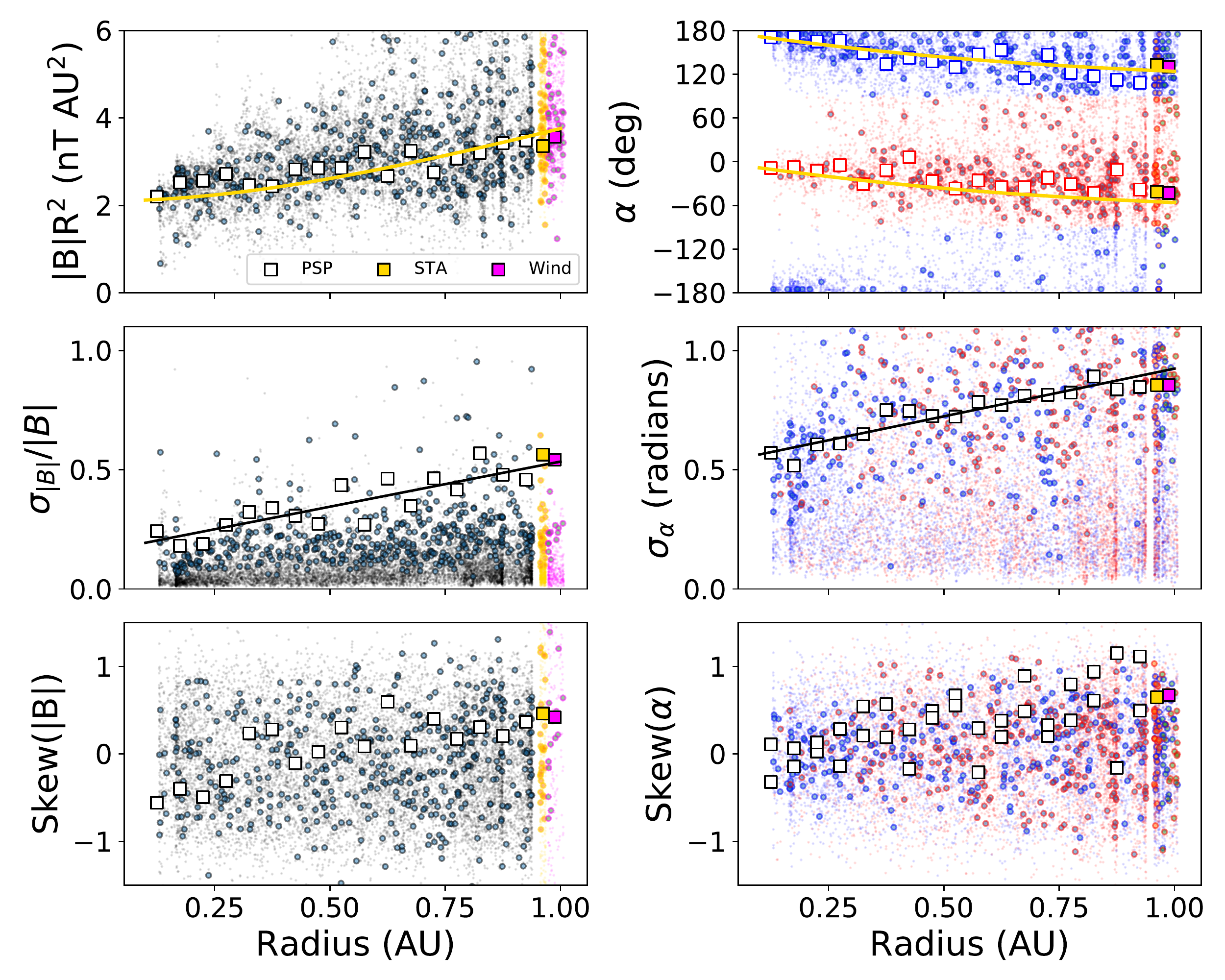}
    \caption{Field magnitude and clock angle statistics as a function of radius. In each panel, the faint background shows statistics of one hour intervals, the larger, darker scatter points show statistics of one-day intervals, and the squares depict the statistics within radial bins of width 0.05AU. The gold and the magenta square depict the STA and Wind results, respectively. The left-hand column pertains to $|B| R^2$ (the field magnitude) and the right-hand column to $\alpha$ (clock angle). The top row depicts the measured large-scale structure and shows the most probable values. In yellow, we have the expectation of a Parker spiral model. The middle row shows the dimensionless standard deviations and a black line shows a least squares fitted linear trend, which describes the radial evolution of vector fluctuations. The bottom row shows the skew of the ratio of the (mean-mode)/standard deviation, which describes the extent to which individual distributions are evenly distributed. For the angular quantities, the red and blue colours describe anti-sunward and sunward sectors, respectively.
}
    \label{fig:parker_spiral}
\end{figure*}

In Section \ref{sec:synthetic}, we found that for idealised synthetic data with symmetrically distributed 2D vector fluctuations, the nature of these fluctuations and the large-scale inclination of the Parker spiral could affect our estimates of the heliospheric flux. In particular, how the fluctuations were partitioned into angular and compressive (|B|) fluctuations had an impact. In this section, we seek to measure the inclination of the Parker spiral and the distribution of 2D fluctuations to assess how these effects manifest in estimates of the heliospheric flux with real data.

In Figure \ref{fig:parker_spiral}, we compute the statistics of the field magnitude (scaled by $R^2$) and clock angle and then we plot the results as a function of radius. We do this by binning all the data by radius (large squares) as well as by binning in time at a daily and hourly cadence (faint and fainter circles). The top row shows the most probable value, the second row shows the standard deviation (for the field magnitude, scaled by the mode to make it dimensionless), and the bottom row shows the skew, defined as the (mean - mode)/standard deviation. 

In the top row, we also plot the expectation of these quantities for a Parker spiral model as a function of radius. The analytical expressions \citep{Parker1958} are :

\begin{align*}
\centering
|B|R^2 = |B_0| \sqrt{1 + \bigg(\frac{R\Omega_\odot}{v_{SW}}\bigg)^2}, \hspace{5mm} \alpha_\pm = atan2\bigg(\mp \frac{R\Omega_\odot}{v_{SW}},\pm1 \bigg).
\end{align*}

Here, $v_{SW}$ is the solar wind velocity, $\Omega_\odot$ is the solar equatorial rotation rate (14.17 deg/day), and $+(-)$ indicates the anti-sunward (sunward) sectors. The models shown use $B_0 = 2.2nT$ and $v_{SW} = 300 km/s$. As can be seen, in all binning schemes, the most probable value of the data is in good agreement with the model across different radii, with larger scatter for the shorter binning intervals.

For the standard deviations (middle row, Figure \ref{fig:parker_spiral}), which describe the relative strength of fluctuations, we see both compressive and rotational fluctuations grow in amplitude with radius, but we also see that at all radii, the amplitude of rotational fluctuations is larger than compressive fluctuations. Binned by radii (such that the distribution is formed over many streams), rotational fluctuations are about three times as large as the compressive fluctuations at closest approach and slightly less than two times as large at 1AU.

The pattern in the skew is less clear. At a one-hour or one-day cadence, there is no trend: at all radii, there are as many distributions skewed positive as negative. For vector magnitude, averaged over many streams, there is a hint of negative skew at perihelion and a small positive skew at 1AU. The negative skew at low heliocentric distances arises from the near-HCS weaker field mentioned in Figure \ref{fig:br_vs_radius}; many single-day intervals at these radii have very low skew. For the clock angle, the skew at perihelion is negligible, but trends positive at 1AU, which means the fluctuations are biased in favour of rotation towards the radial direction.

The relationship of the different timescales considered here and the underlying data products warrants some discussion in passing. Firstly, we note explicitly that the one-hour and one-day cadence data points are computed from the higher cadence four-samples-per-cycle data product, while the data binned by radius (white squares) are based on the 1 min averaged data. The fact that the white squares lie in the middle of the distribution of one-hour and one-day measurements fuels our confidence that the two data products are largely interchangeable. This also shows that at a given radius, non-consecutive hour or day intervals from different orbits with no temporal correlation fill out a well-defined distribution around the relevant white square. This is useful evidence that the central value is not tied to a specific time interval and it is, therefore, real spatial information about the HMF, meaning it is justified to use different timescales for estimating the heliospheric flux at various points in this paper. 

To summarise, in this section, we successfully measured the 2D vector fluctuations and underlying large-scale radial trends in the HMF that, as we showed in Section \ref{sec:synthetic}, could impact our measurements of the heliospheric flux. In addition, we characterised the skew of the distributions, broadly assessing the requirement for the vector fluctuations to be evenly distributed about a central value to justify the use of the Parker spiral method.  Additionally, based on this, we infer  that: (1) the Parker spiral is a good model for the underlying large-scale structure; (2) at all radii from 0.1-1AU, the amplitude of rotational fluctuations are stronger than compressive ones ($\sigma_{|B|}/|B| < \sigma_\alpha)$,  but this predominance decreases with heliocentric distance; (3) the Parker spiral method for computing $\Phi_H$ (which relies on having an unskewed distribution in magnitude and clock angle) may be less applicable at 1AU, where the distributions averaged over streams appear to have a positively biased skew in both polar components (and the central assumption of evenly distributed fluctuations breaks down). The field magnitude over multiple orbits also showed some signature of skew close to the sun, but this can be attributed to distortion by the weaker field near the HCS during encounters four and five. This is accounted for by taking the most probable value of the magnitude rather than the mean. For reference, the radially binned distributions which give rise to the statistics shown in this section are shown in Appendix \ref{sec:appendix:synth_vs_real}. Based on Figure \ref{fig:synthetic} and the above observation that the rotational fluctuations are dominant, we expect the mean to be a systematic underestimation of the background value of $B_R R^2$, especially close to the sun. 

\subsection{Synthetic and measured radial trends of flux}
\label{sec:measured_trends}

\begin{figure*}[h!]
    \centering
    \includegraphics[width=0.92\textwidth]{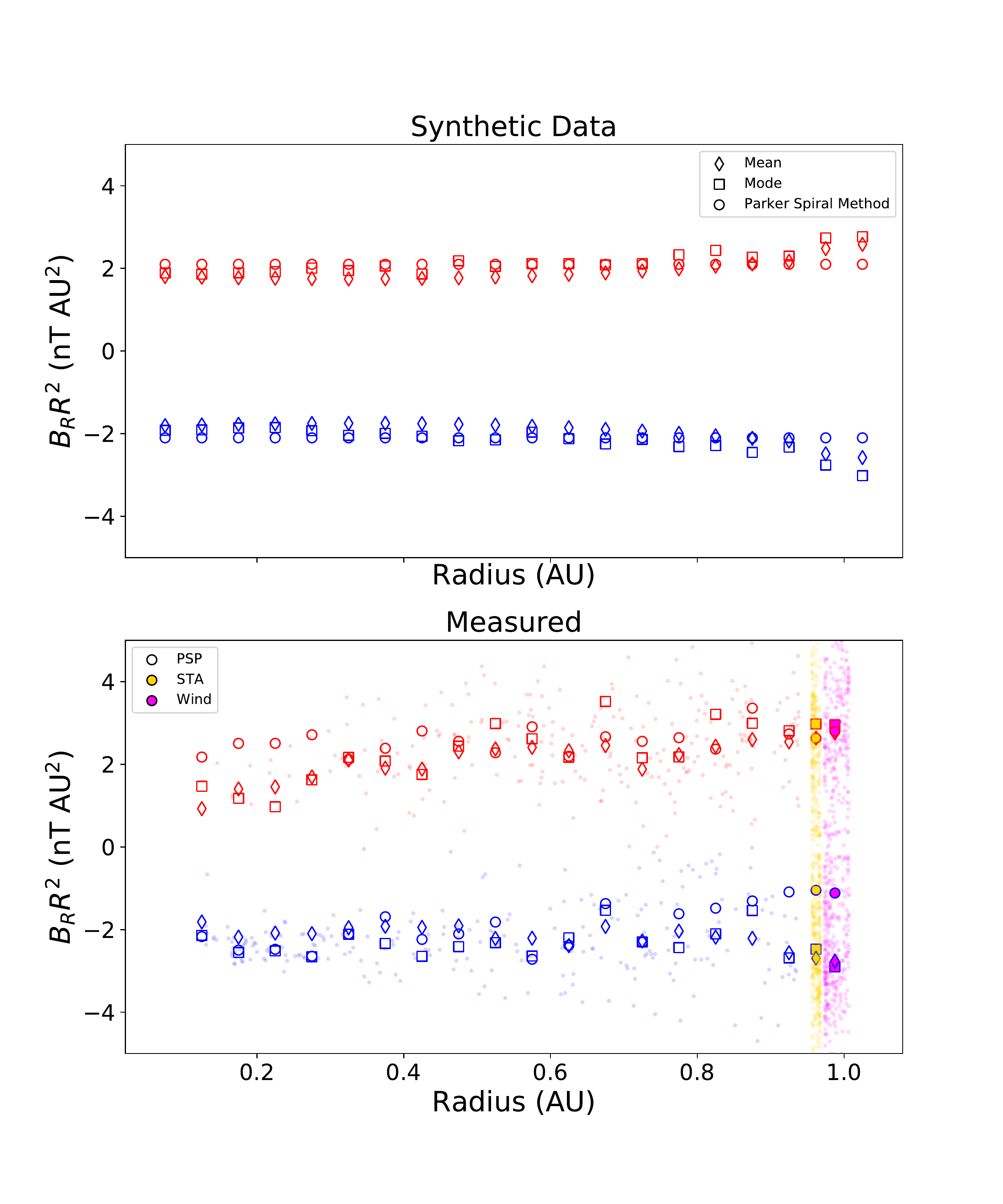}
    \vspace{-15mm}
    \caption{
    Resulting expectations and measurements of the difference between the different measures of $B_R R^2$. In both panels, circles indicate results from the Parker spiral method, squares represent modes, and diamonds represent means, as a function of radius. The top panel shows results for a purely synthetic data set using a Parker model for the “true” field and Gaussian* fluctuations in magnitude and clock angle as modelled by the linear trends shown in fig \ref{fig:parker_spiral}. The bottom panel shows the same measurements from the actual PSP data, binned into 0.05AU radial bins. Faint data points in the background show the computed values at one day intervals to show the general scatter (which clearly increases with radius --  except where Gaussian fluctuations would cause negative values of |B|; for details, see main text).
}
    \label{fig:flux_vs_radius}
\end{figure*}

Having measured the large scale variation of the HMF and the characteristics of the vector fluctuations, we next seek to quantify the expected and measured variation in the statistical methods of estimating $B_R R^2$ as a function of radius. In Figure \ref{fig:flux_vs_radius}, we compare measurements of the mean (diamonds), modes (squares), and the Parker spiral method (circles) from purely synthetic distributions (top panel) and directly from measurements (bottom panel).

For the synthetic measurements, we take the Parker spiral model values for the background values of $|B_0|(R)$ and $\alpha_0(R)$ from Figure \ref{fig:parker_spiral} (yellow curves, top panels) and for the fluctuation amplitudes, we take the linear fits to the standard deviations from the middle panels of Figure \ref{fig:parker_spiral}. As earlier, even though we compute the two sectors separately for the synthetic data, we use the joint distribution projected into 1D and split at $B_R R^2 = 0$ to compute the mean and mode. This method simulates the process of obtaining these measurements from the real data where the components of the two magnetic sectors cannot be distinguished in the 1D distribution. We do actually allow the synthetic distribution in magnitudes to have a non-zero skew further from the sun (see Appendix \ref{sec:appendix:synth_vs_real} since fitting a standard deviation from these skewed distributions would produce an nonphysical Gaussian distribution, where a significant population of magnitude values would go through zero. In these cases, we are modelling our `ground truth' as the peak of the skewed distribution of magnitude. 

As expected, for the purely synthetic data, the Parker spiral method produces the same value of magnetic flux for all radii. While this background value remains constant, the estimated mean and modes evolve with radius due to the changing impact of the fluctuations and background Parker spiral angle. For low radii, the mean produces a constant value but significantly underestimates the magnetic flux. Around 1AU, the mean systematically increases relative to the background value, and in this model becomes an overestimation. Closer than about 0.8 AU, the mode is a slightly better estimate of the background value but it fluctuates, at some points, resulting in either an overestimation -- or sometimes an underestimation. At 1AU, the mode also systematically increases and overestimates the `true' value.

For the real data, the PSM is performed directly with the most probable values of |B| and $\alpha$ taken from the white squares in Figure \ref{fig:parker_spiral} to show representative values integrated across multiple orbits. These are derived with the 1 min averaged data products. We note that the yellow and magenta filled data points are from STEREO A and Wind respectively, while the remainder of the data is from PSP. For completeness, we show both the positive and negative sectors measured by PSP, although the trends discussed below are primarily observed in the negative polarity (blue data points). This is due to the limited time intervals where PSP has been above the HCS, particularly at the closest approach. Furthermore, the dominant data contribution above the HCS for radii closer than 0.3AU comes from encounters four and five, which were distorted for extended periods by PSP skirting very close to the HCS (Chen et al., submitted to this issue). 

As faint data points in the background, PSM estimates of $B_R R^2$ for one day time intervals are shown from the four-samples-per-cycle data, demonstrating the variability of the estimate from one day to another and that the values of the estimate integrated over multiple orbits lie in the centre of this distribution. This distribution is more spread than the difference in the central value from one radial bin to the next.

Although the trends with real data are unavoidably noisier, when focusing on the negative sector, the mean can be clearly seen to exhibit the same behaviour as in the synthetic data, producing a systematic underestimation close to the Sun, but growing and exceeding the Parker spiral method at 1AU. In the positive sector, the same behaviour is seen, but less clearly for R > 0.3AU, with a flat mean eventually trending upwards for R > 0.8AU. Importantly, the trends implied by the PSP data converge nicely with the 1AU data. For the negative polarity data trends the mode is also in good agreement with the synthetic observations, providing better agreement with the Parker spiral method close to the Sun, but also systematically increasing near 1AU and at intermediate radii overestimating the PSM. The PSM, while not perfectly flat, is generally quite consistent with the radius, but not dramatically more so than the other methods. For $B_R R^2 < 0$ near 1AU, there is a significant drop-off in the estimate of $B_R$ using the Parker spiral method. This may be related to the growing skew exhibited in the bottom panel of Figure \ref{fig:parker_spiral} and suggests the method may be less robust near 1AU. As mentioned previously, our implementation of the PSM relies on the fluctuations (particularly in the angle) being evenly distributed about a mean. The bottom panels depicting the skew in Figure \ref{fig:parker_spiral} suggests this assumption starts to breaks down close to 1AU.

Overall, we conclude that this model of fluctuations in magnitude and angle is a good approximation for understanding the dependence on heliocentric distance of the different methods of computing $\Phi_H$ considered here, at least for R < 0.8 AU. In particular, Figure \ref{fig:flux_vs_radius} suggests that fluctuations cause the mean of PSP data close to the Sun to constitute an underestimation -- and to be an overestimation beyond 0.8AU. Below 0.8AU, at least for the negative polarity measurements, the PSM and mode estimates are comparably variable with radius but show the ordering expected from our model of fluctuations -- with a higher value of mode as compared to PSM and a convergence of the values close to the sun. Comparisons of the synthetic and measured 2D distributions and distributions of $|B|$ and $\alpha$ in 0.1AU bins may be found for reference in Appendix \ref{sec:appendix:synth_vs_real}. 

As a result of the findings in this section, we conclude that it is necessary and important to utilise the 2D distribution of the magnetic flux in order to compute the background value of $B_R R^2$; without this step, the radius-dependent fluctuations can lead to systematic biases in the mean or modal statistics value. Thus, of the methods considered here, the Parker spiral method presents the most robust estimate, although its estimate appears interchangeable with the mode close to the sun and may be less reliable near 1AU. In the next section, we present our results, using the Parker spiral method (and the other statistical methods where justified) to measure $B_R R^2$ throughout the inner heliosphere and, finally, we compare it to expectations from coronal models.

\section{Results}
\label{sec:results}

\begin{figure*}[h!]
    \centering
    \includegraphics[width=0.9\textwidth]{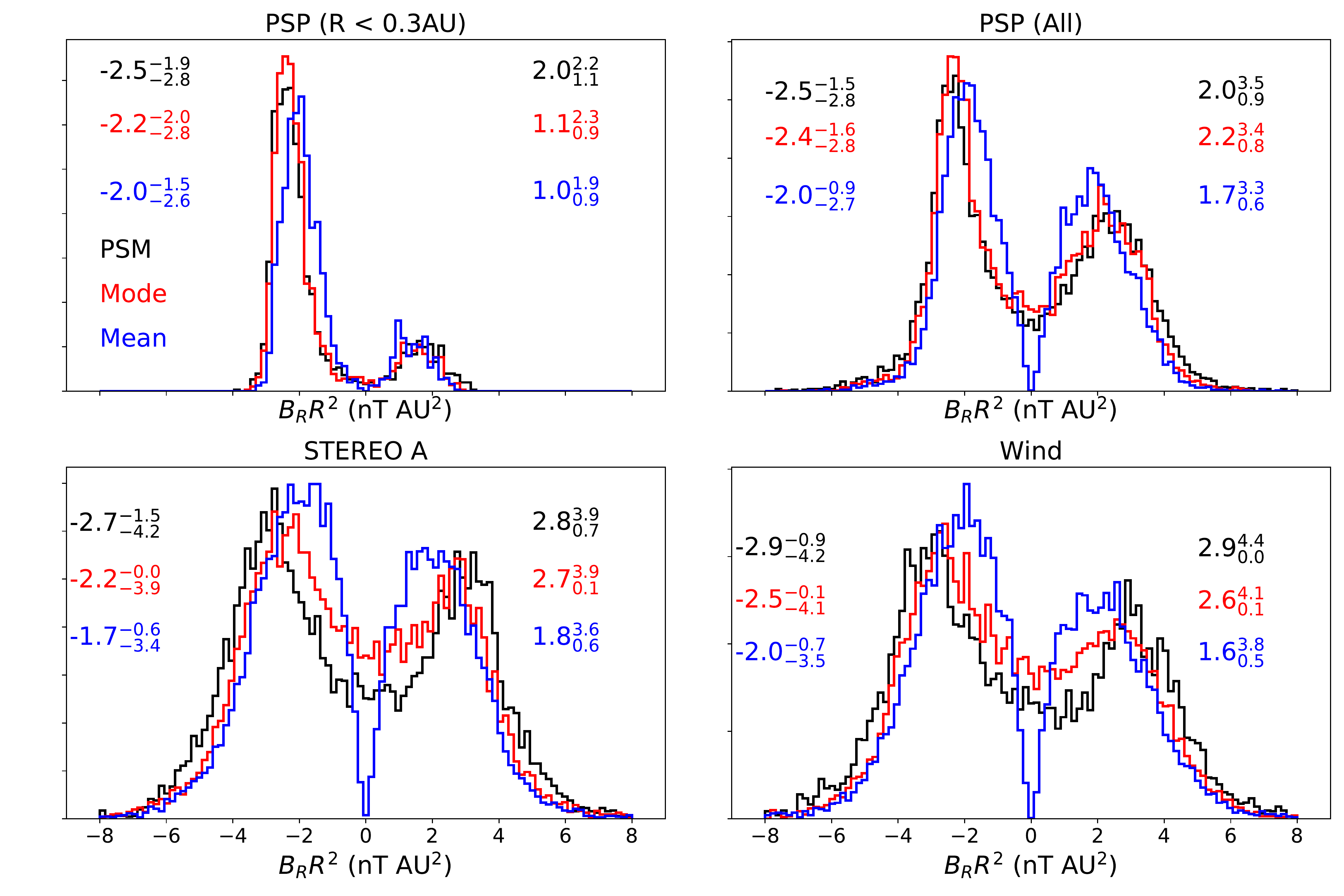}
    \caption{Bulk statistics. Each panel shows 1D histograms of all values of $B_R R^2$ computed with the Parker spiral method (black), mode (red), and mean (blue) at one-hour timescales, integrated over all the data shown in Figure \ref{fig:synoptic}. Reading from left to right, the four individual panels  show PSP (for radii less than 0.3 AU), PSP (all data), STA and Wind data respectively. Text in the corresponding colour gives the bi-modal peak values and their upper and lower values defining their full width at half maximum (FWHM) in superscript and subscript.
}
    \label{fig:bulk_measurement}
\end{figure*}

\subsection{Bulk measurements}
\label{sec:bulk}

In Figure \ref{fig:flux_vs_radius}, near 1AU, at least for the positive polarity data, all three metrics appear to climb to near 3nT AU$^2$, possibly indicating a physical enhancement in the heliospheric flux at 1AU. To probe this observation further, as well as to develop a `best estimate' of the heliospheric flux closer to the sun, we sum together all the $B_R R^2$ distributions over radius and show their bulk statistics. For every hour of data, we compute the mean, mode, and Parker spiral method estimates as described in Section \ref{sec:2d_vs_r} and show how these estimates are distributed over the studied time interval. A one-hour timescale is chosen here to produce a sufficient number of individual `measurements (on the order of 15000) with each method across the whole data set expected to produce a robust histogram and define a peak value. A longer integration timescale would sharpen these distributions, but  limit the statistics. This means our computations of error are likely conservative.

The results are shown in Figure \ref{fig:bulk_measurement}. We show the three spacecraft separately -- for PSP we include a panel showing only data from close to the sun (R < 0.3 AU) -- and overlay the histograms of hourly measurements with the three different methods. The information contained in these histograms is compressed into a most probable value and upper and lower values defining the full width half maximum (FWHM) for both peaks. These data are shown as text in the corresponding colour in the same plot. 

We first observe that in all cases, the distributions of measurements using the mode and Parker spiral methods are quite similar to each other, whereas the distribution of mean measurements is strongly distorted: while the truncation procedure (similar to taking the modulus, see Section \ref{sec:2d_vs_r}) successfully keeps the sunward and anti-sunward peaks distinguishable, the overall distribution is broadened and shifted towards $B_R R^2 = 0$, producing a systematic underestimation compared to the other methods and with a larger uncertainty as encoded by the FWHM.  We see by restricting the data to PSP measurements close to the Sun  (top left panel in Figure \ref{fig:bulk_measurement}), the distributions for all methods are the sharpest, although the systematic underestimation between the mean and the other methods persist, as expected from our synthetic data model. The broadening of all distributions with distance from the sun shows the increasingly prominent role of fluctuations and other distorting effects with increasing heliocentric distance.

Distinguishing the mode and PSM is slightly more subtle. With regard to the near-1AU distributions, the Parker spiral method is seen to produce distributions that are more similar to the close approach distributions further from the Sun, doing a better job of removing the population of measurements close to zero and separating the two peaks, as compared to the mode. While this is most clearly seen at 1AU where there is a larger population of $B_R R^2$ measurements close to zero in the red curve (mode) compared to the black curve (PSM), integrating the PSP data over all radii, (top right panel in Figure \ref{fig:bulk_measurement}), the difference is slight but still observable.  Once we restrict the data below 0.3 AU (top left panel in Figure \ref{fig:bulk_measurement}) both the mode and PSM distributions are close to indistinguishable. This supports our inference, presented in Section \ref{sec:measured_trends}, that the mode is distorted more than the PSM by fluctuations as distance from the sun grows, although for practical purposes, for most of the PSP data, the noise in these distributions means the mode and PSM measurements derived are consistent within the error; the growth of the difference between these methods is most pronounced between the outer ranges of PSP's orbit and 1AU.

We note for the PSP data, the negative peaks are much better resolved than the positive peak due to the same latitudinal sampling effects and near-HCS intervals which caused the trends to be more apparent with the negative polarity data in Figure \ref{fig:flux_vs_radius}. Integrated over all radii, the negative polarity PSM peak gives a bulk value of $\Phi_H = 2.5^{+0.3}_{-1.0}$nT $AU^2$ where, as mentioned above, we have computed the error based on the FWHM of this profile. Confined to just R < 0.3 AU, the measurement tightens to $\Phi_H = 2.5^{+0.3}_{-0.6}$nT AU$^2$.

We also see the most probable PSM-derived estimate of $B_R R^2$ at 1AU is significantly larger ($\sim$2.8 nT $AU^2$) than the value derived above from the negative peak of the PSP data, although just within the error. The enhancement is apparent in both Wind and STEREO A data. The trend is also seen in the most probable value of modal values, but not here for the mean. For the mean, this seems to be an issue with defining the mode for a broad peak, looking at the upper FWHM value, this can be seen to clearly increase for all methods in comparing the top two panels to the bottom two. While we expected from the analysis of fluctuations that the mean and mode of $B_R R^2$ will grow with radius, this cannot explain why the value would be significantly higher than computed with the Parker spiral method. Although we expect the PSM results to be less robust in this vicinity due to apparent skew of the distributions and associated breakdown of the even distribution assumption (Section \ref{sec:measured_flucs}), it is not immediately apparent why this should result in a systematic increase of $B_R R^2$ as opposed to a decrease. Therefore, it appears possible that this flux enhancement could indeed be physical. We investigate a possible explanation for such an enhancement in the next section.

\subsection{Excess flux}
\label{sec:excess_flux}

\begin{figure*}[h!]
    \centering
    \includegraphics[width=0.9\textwidth]{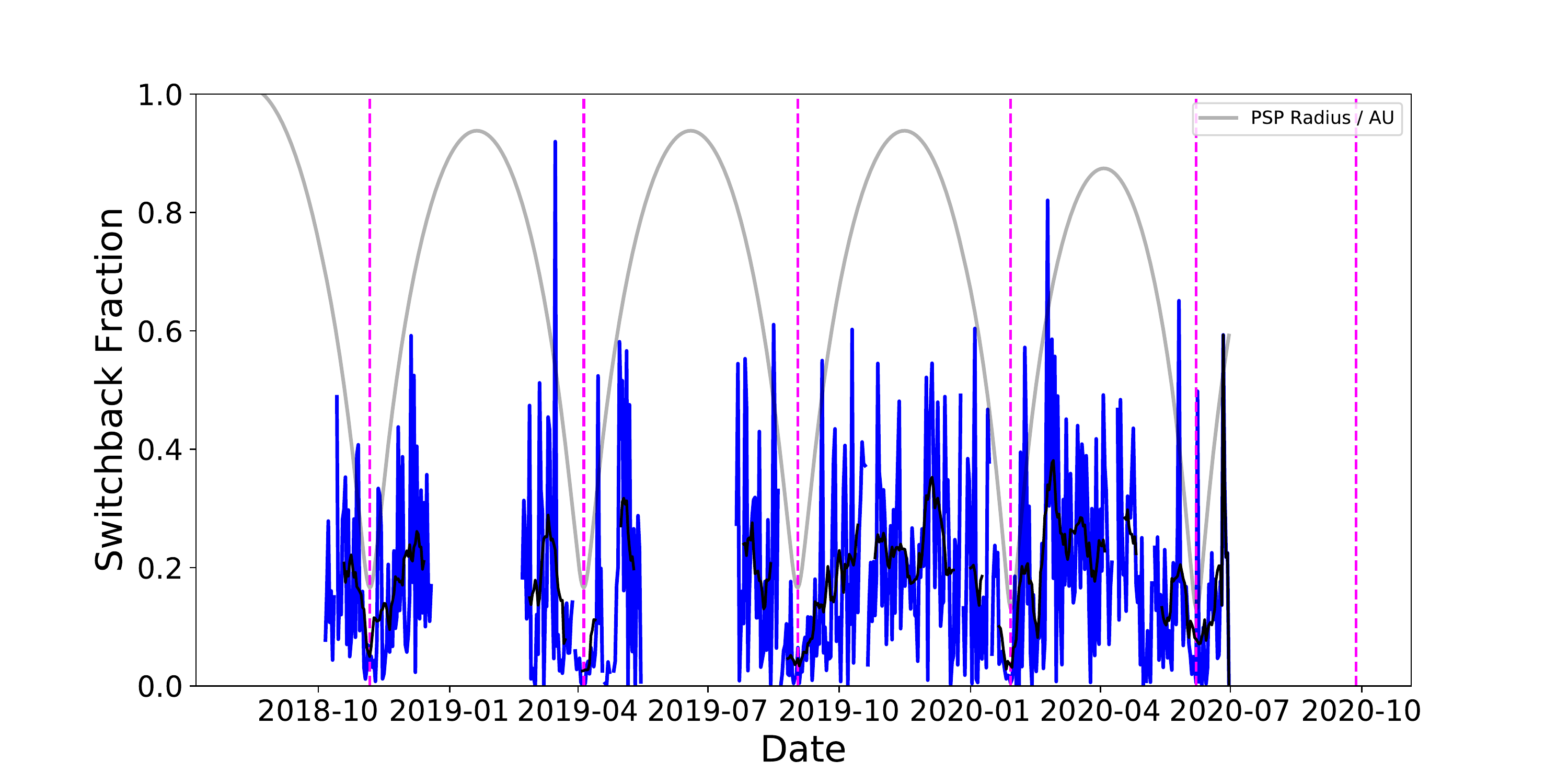}
    \caption{ Daily switchback fraction. The blue curve shows, for each day of PSP data, the fraction of measured $B_R$ values which are of the opposite sign to the most probable field that day. The light grey curve show’s PSP radius in units of AU. The black curve better illustrates the underlying trend, showing the daily fractions smoothed with a window size of five days. The dashed magenta lines indicate the date of each perihelion.
}
    \label{fig:switchback_fraction}
\end{figure*}

In the heliosphere, numerous physical processes such as waves, turbulence and stream interactions perturb the quiescent picture implied by the Parker model. One particular consequence of such effects is that local field inversions develop wherein the field lines connecting back to the corona warp into an S shape. As argued by \citet{Owens2017}, such inversions entail the same field line threading a spherical surface at a fixed radius multiple times (see also Appendix \ref{fig:appendix:sb_schem} for an illustration) such that a given field line would contribute triple the flux at this radius as compared to its contribution leaving the corona. Therefore, they could contribute to a physical excess flux. The cited authors identify such inversions using the electron heat flux (aka strahl) which follows the HMF topology and  carries energy away from that field line's point of origin in the corona. Therefore, when the strahl is observed to be sunward, it is inferred that the electrons must be travelling along a kinked field line such that they escaped and moved outwards from the sun but have then been guided sunward by a field inversion. This approach was recently applied  \citep{Kasper2019} to probe the topology of the switchbacks observed by PSP. \citet{Owens2017} used this method with ACE data over a solar cycle to derive a correction factor to the heliospheric flux. They found this estimate to be consistent with the values obtained by pre-averaging signed $B_R R^2$ data over one day intervals before taking the modulus, and slightly weaker than the kinematic correction factor proposed by \cite{Lockwood2009a,Lockwood2009b,Lockwood2009c}. Furthermore, \citet{Macneil2020} showed with Helios data that the frequency of such local inversions increased with radius from 0.3 to 1.0 AU, suggesting that this correction factor would grow with radius.

In Figure \ref{fig:switchback_fraction}, we investigate this same radial trend with PSP $B_R$ data from 0.13 to 0.9 AU. For each day of PSP data, we compute the fraction of the distribution which is of the opposing sign to the most probable value, making the assumption that over each 24-hour period, PSP remains on one side of the HCS and that all zero crossings in this interval are due to fluctuations. This is slightly less robust than the \citet{Owens2017} strahl method which can distinguish HCS crossings unambiguously and account for transient structures such as coronal mass ejections (CMEs) but still carries useful information about the trend. A one-day interval is chosen as the trend is readily apparent with no further processing; computing it with a shorter time interval, such as one hour, is much noisier. In Figure \ref{fig:switchback_fraction}, we plot this fraction of inverted flux as a function of time and also show the times of PSP's perihelia and its radial variation. While the metric is noisy, as might be expected given the above assumptions, there is a clear correlation with radius, in which the fraction of inverted flux grows with distance from the Sun, which is consistent with \citet{Macneil2020}. The fraction has a well-defined floor which varies from $<3\%$ at perihelion to as much as 20\% at 1AU. It is therefore plausible that if such fluctuations can contribute to excess flux, their impact will grow with radius in the inner heliosphere. However, conversely, we see that at perihelion the contribution is extremely small (< 3\%). This is interesting given the prominence of the switchbacks observed by PSP. This implies that while these switchbacks are very striking given the large amplitude and sharpness of the rotations \citep[e.g.  ][]{Horbury2020}, the population of measurements in which the field actually reverses is a very small fraction of the total. They are transient impacts perturbing a quiescent background state. 

To numerically estimate the possible impact of the flux, we make a very simple construction, which is as follows: if there are $N$ open field lines emerging from the corona and a fraction $f$ of them are locally inverted at a given radius, then the true open flux per unit area is $d\Phi_{open}/dA = N; $ but the actual measured flux at that radius will be $ d \Phi_H / dA_m = (1 - f)N + 3fN$. This can be rearranged to relate the true and measured flux : $\Phi_{open} = 1/(1+2f) \Phi_{H}$. For $f=3\%$, this simple heuristic implies $< \%5$ open flux estimate error at perihelion and up to a $30\%$ correction at aphelion, which is similar (slightly larger) than the factor derived by \citet{Owens2017}, as judged from their Figure 5, and significantly larger than the fraction implied by Figure \ref{fig:bulk_measurement} in the present work, which suggests a $\sim 20\%$ reduction in flux between PSP and 1AU. This metric is likely an overestimation since it assumes the switchback fraction measured near the ecliptic plane, where the solar wind is predominantly slow, can be applied uniformly in latitude but, as shown with Ulysses data \citep{Erdos2012}, fast polar coronal hole wind for most latitudes has much lower fluctuation levels than the slow wind. 

Regardless, in terms of contribution to the excess flux, near PSP's perihelia flux inversions are largely negligible. Combined with the much reduced relative amplitude of fluctuations at closest approach, we expect that the value of heliospheric flux, $\Phi_H$, at least at PSP's closest approach, should robustly correspond to the true open flux escaping the corona, $\Phi_{open}$, with $5\%$ as an upper bound to the possible deviation.

\subsection{Heliospheric flux as a function of heliographic position}
\label{sec:flux_vs_loc}

\begin{figure*}[h!]
    \centering
    \includegraphics[width=\textwidth]{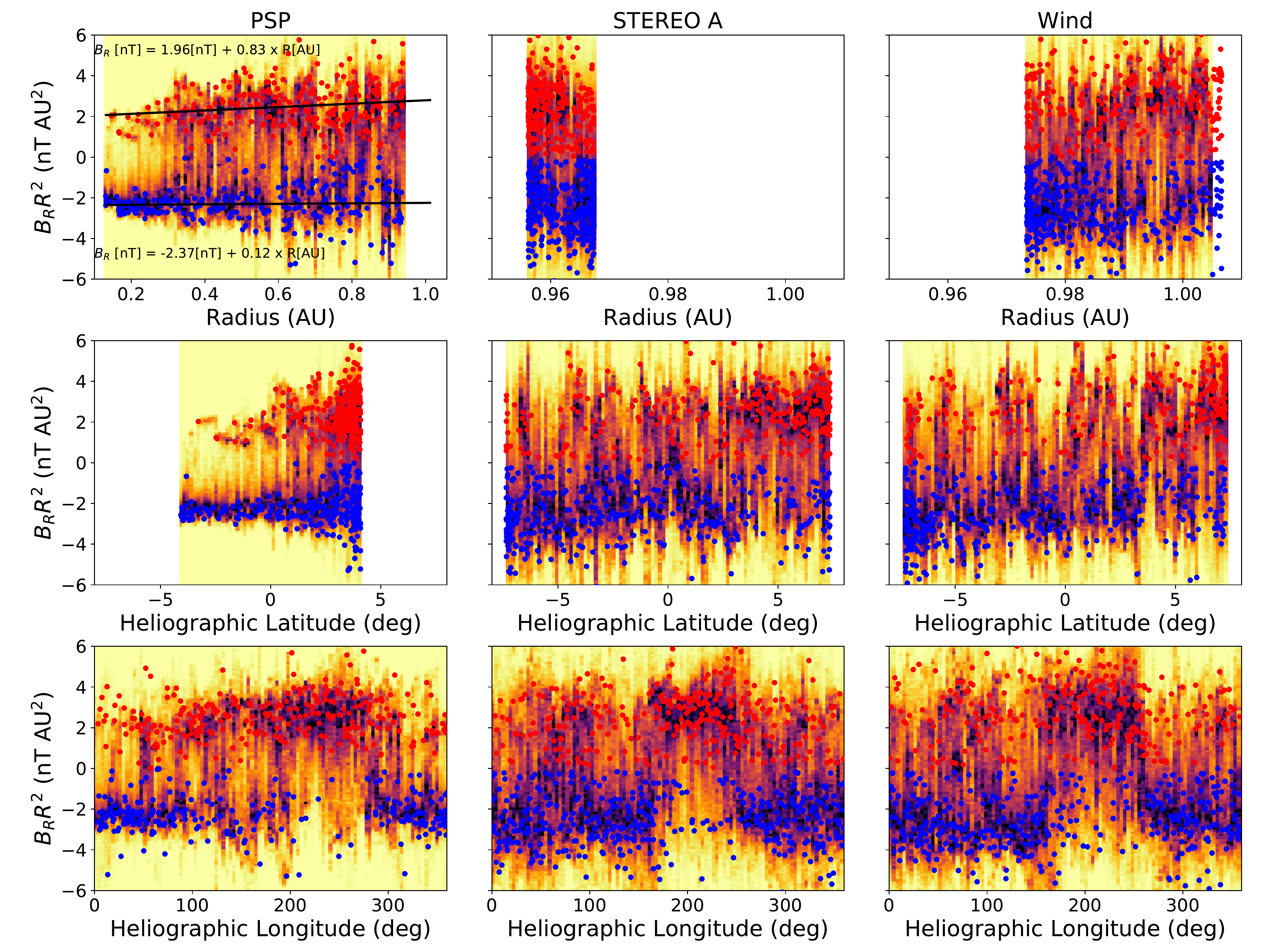}
    \caption{Measured flux as a function of heliographic location. Each panel shows a 2D histogram in the background as in figure 1. In the foreground, blue and red dots show the values computed at one day intervals with the Parker spiral method. The top row shows trends versus radius, the middle offers a comparison against heliographic latitude, and the bottom against heliographic longitude. The middle and bottom rows have a common x-axis. The three columns depict data from PSP, STEREO A and Wind respectively. For the PSP heliospheric flux as a function of radius (top-left panel), solid black lines indicate fitted linear trends to the scatter points shown with negative and positive (blue and red) values fitted separately}.
    \label{fig:flux_vs_loc}
\end{figure*}

We show here how the Parker spiral method determination of heliospheric flux varies with heliographic position, while examining all three dimensions of radius as well as heliographic latitude and longitude. The results are summarised in Figure \ref{fig:flux_vs_loc}. In the background, we show 2D histograms of the full data set of 1 min averages (again due to tractability issues of working with the full four-samples-per-cycle data set). In the foreground, we plot the Parker spiral method results computed from each day of data and colour them by the polarity. One-day intervals are used to reduce the scatter and not overcrowd the plot as opposed to one-hour intervals, while spatial binning is avoided so that each data point may simply be assigned a 3D spatial coordinate so that it may easily be plotted against each coordinate (radius, latitude, and longitude). The top row shows results against radius, the middle against latitude, and the bottom row against longitude. Columns from left to right shows PSP, STEREO A and Wind results. 

First we see that in all cases, the one-hour estimates trace the regions of highest data density (dark colours in the histograms) well, indicating the close relationship between the Parker spiral method and the mode. Second, plotted against all the coordinates, the PSP data shows the tightest confinement and least scatter due to the much reduced fluctuation levels (see Figure \ref{fig:parker_spiral}, middle row). We confirm the heliospheric flux over these two year intervals is largely independent of radius and longitude, but most clearly shown by the PSP data for $R$ < 0.5AU and for negative latitudes. To emphasise the radial trend, we make linear fits to the data shown (with negative and positive data points fitted separately). These fits are shown as solid black lines in the top left panel of figure \ref{fig:flux_vs_loc} and their fitted equations are printed within the same panel. The fit to positive polarity shows a substantial slope due to the positive polarity of the weakly magnetised plasma PSP sampled at it running close to the HCS at closest approach during encounters 4 and 5 (see Figure \ref{fig:br_vs_radius} and Chen et al., submitted to this issue). The negative polarity data is much less perturbed by these effects and benefits from a larger sample size (three whole encounters predominantly below the HCS). For this case, the trend is well fit by a constant value indicating constant heliospheric flux as a function of radius and limiting the variation in the flux with radius to 0.1 nT AU$^2$ per AU. In fact, this fit suggests a weak decrease in flux with radius. We note that as suggested by Figure \ref{fig:flux_vs_radius}, the onset of increasing flux with distance may be non-linear and may inflect close to 1AU.

Over the limited range of latitudes probed by PSP and the 1AU spacecraft, the heliospheric flux is also seen to be constant. It is no coincidence that the latitude and radius panels show very similar trends since these coordinates are strongly correlated due to the tilt of the spacecraft orbits relative to the solar equatorial plane. The high data scatter at high PSP latitudes is simply a result of PSP being at aphelion during the sampling at these high latitudes, where the fluctuation environment is much noisier.

Comparing the longitude plots, we see a similar sector structure pattern measured by all three spacecraft over the full two-year data set, indicating the dominant warps in the HCS were consistent throughout. The predominantly positive sector in the PSP panel (top right) is shifted to a slightly higher longitude than on the 1AU panels, which is consistent with the Parker spiral shift from PSP's closer heliocentric distance out to 1AU. 

The STA and Wind data show a slight asymmetry in field scatter by latitude. At highest latitude, they measure slightly higher positive flux, and at lowest latitude they measure slightly stronger negative flux. The reason is not perfectly clear. From Figure \ref{fig:synoptic}, it is clear that the latitude and field values are correlated. Extremes in latitude means the spacecraft is spending less time close to the HCS in general. This might mean faster solar wind with lower fluctuation levels and therefore a measured value which is closer to the 'true' open flux and would suggest the excess flux is physical. 

\subsection{Heliospheric flux as a function of time and comparison to PFSS expectations}
\label{sec:flux_vs_time}

\begin{figure*}[h!]
    \centering
    \includegraphics[width=\textwidth]{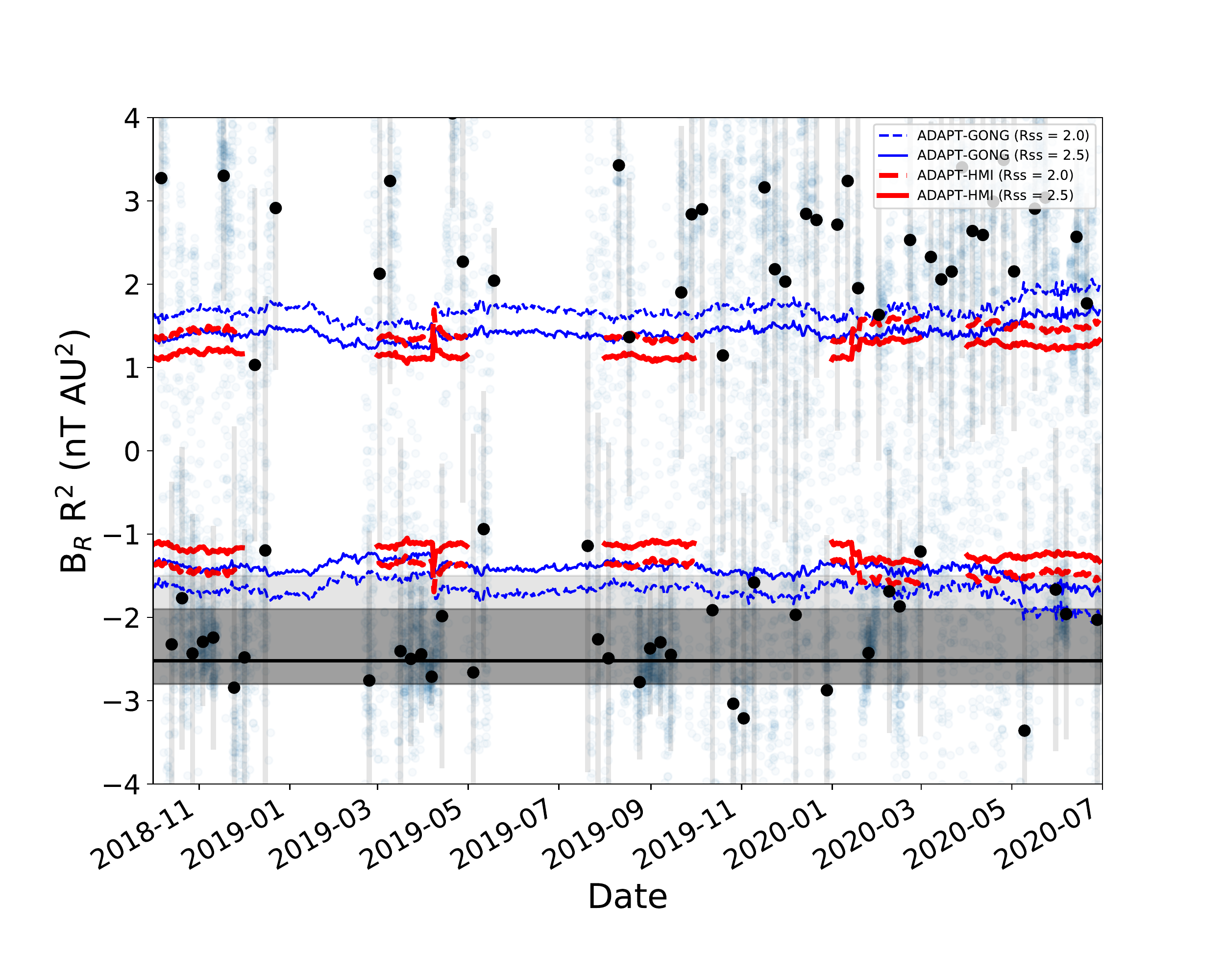}
    \caption{$B_R R^2$ versus time and Potential Field Source Surface Flux estimates. As a function of time, black markers represent the most probable Parker spiral method value (at a one hour cadence) of the \textit{in situ} heliospheric flux for each week of data, the grey error bars represent the standard deviation with this week. The light gray and darker grey regions represent the full width half maxima of all PSM method measurements and all measurements within 0.3AU respectively. A horizontal black line shows the most probable value (0.25 nT AU$^2$) which is common to both the full and radius-restricted data sets (see figure 7). Blue (ADAPT-GONG) and red (ADAPT-HMI) curves show the open flux value predicted (see Equation \ref{eqn:ss_method}) by potential source surface (PFSS) models using daily updating magnetograms over the mission. The solid (dashed) curves show results for 2.5 $R_\odot$ ($2.0 R_\odot$) source surface height, and the different colours differentiate the magnetogram source used.
}
    \label{fig:flux_vs_time_pfss}
\end{figure*}

Lastly, we compare the Parker spiral method measurement results to estimates of the open flux from Potential Field Source Surface models \citep[PFSS][]{Altschuler1969,Schatten1969,Wang1992}. The specific model implementation is the same as described in \citet{Badman2020} and makes use of \textit{pfsspy} \citep{Yeates2018,Stansby2019a,Stansby2020}, a python implementation of the PFSS model. The inputs of the model are 2D magnetogram maps of $B_R$ at the solar photosphere, and some choice of source surface height ($R_{SS}$). The output of the model is the vector magnetic field, $\boldsymbol{B(\boldsymbol{r})}$, for the annular region between the photosphere and the source surface ($1R_\odot < R < R_{SS}$). The open flux is estimated from this model by integrating the field vector over the source surface, where all field lines are radial by construction and interpreted to be open to the solar wind. This integral is given by Equation \ref{eqn:ss_method}, which is repeated here for convenience: :

\begin{align*}
    \Phi_{open} = \int_0^{2\pi} \int_{-\frac{\pi}{2}}^{\frac{\pi}{2}} |B_R(\theta,\phi,R=R_{SS})| R_{SS}^2 \sin\theta d\theta d\phi.
    \label{eqn:ss_method_again}
\end{align*}

As a reminder, to make this quantity comparable to the quantity $B_R R^2$ measured in situ, we normalise it by $4\pi$ (1AU)$^2$, such that is is expressed as a field strength in nT at 1AU.

We used the Air Force Data Assimilative Photospheric Flux Transport \citep[ADAPT, ][]{Arge2010} magnetograms \footnote{ADAPT maps are accessed from \url{https://gong.nso.edu/adapt/maps/}.},  which assimilate the most recent available photospheric data (from the visible part of the Sun) into a surface flux transport model which forward-models the magnetogram from the previous time-step, taking into account differential rotation and meridional flux transport. In this way, it seeks to model the magnetic field of the entire solar photosphere `synchronically' (meaning all longitudes are modelled at the same time), as opposed to synoptic magnetograms, which simply merge together old and new data. These maps are produced multiple times per day (much more frequently than a Carrington rotation). In this work, we access one magnetogram per day. We utilise ADAPT magnetograms, which assimilate photospheric data from the ground-based Global Oscillation Network Group \citep[GONG; ][]{Harvey1996} and space-based Heliospheric Magnetic Imager \citep[HMI; ][]{Scherrer2012} which are both available at the above URL for the PSP mission duration. The HMI data product is only available for about $\pm1$ month about each perihelion, while the GONG data product is available for the full duration studied.

In Figure \ref{fig:flux_vs_time_pfss}, we compare the open flux estimated from the PFSS models to that measured in situ by PSP as a function of time. Results for two different values of source surface heights ($2.0R_\odot, 2.5R_\odot$) are shown for ADAPT-GONG and ADAPT-HMI maps. The lower value is chosen based on better agreement with the current sheet crossings measuring during PSP's first perihelion \citep{Badman2020,Reville2020, Szabo2020}, while the higher value is the widely accepted canonical value of $R_{SS}$ \citep{Hoeksema1984}, and which appears a better fit for PSP's second orbit \citep{Panasenco2020}.  

To compare to the in situ measurements, we here present seven day averages of the hourly Parker spiral method results from Figure \ref{fig:flux_vs_loc}. This is chosen to avoid overcrowding the plot and emphasise where the data is most self-consistent and why the most probable trend line is located where it is. Error bars show the standard deviation of the hourly measurements within these seven days.  In addition, the most probable value and error generated from the bulk distribution of the PSM results (histogram peak and upper and lower FWHM, top right panel in Figure \ref{fig:bulk_measurement}) is shown as a solid black horizontal line and light grey region.  In addition, the darker grey region shows the smaller error region of the distribution for R < 0.3 AU, which has the same central value (top-left panel, Figure \ref{fig:bulk_measurement}). This error region captures all the data points for the first three perihelia well. As noted in Section \ref{sec:bulk}, these errors were generated using the distribution of one hour PSM measurements. This short time interval was necessary to produce enough statistics to produce a robust distribution, but as we see in the discussion in Section \ref{sec:measured_flucs}, longer time intervals reduce the scatter further and, thus, the error bounds here are likely to be quite conservative. We see that the most reliable measurement of the open flux occurs southwards of the HCS, around perihelion of the first three orbits (November 2018, April 2019 and September 2019).  This `best estimate' is shown only for negative polarity since PSP has been predominantly south of the HCS during each perihelion pass when it measures these low error values and may be summarised as $\Phi_H = \Phi_{open} = 2.5^{+0.3}_{-0.6}$ nT AU$^2$. We note the $5\%$ error implied by the excess flux discussed in Section \ref{sec:excess_flux} is $0.125$ nT AU$^2$ and is sub-dominant to the statistical uncertainty.

Comparing the PFSS and in situ results, we see regardless of time evolution, distribution time interval, source surface height, or choice of magnetogram, the best estimate of open flux measured by PSP exceeds that expected by PFSS estimates. As is to be expected, lowering the source surface height increases the estimated flux. Continuing below 2.0 $R_\odot$ would eventually enhance the flux sufficiently to match the in situ data, but this is at the expense of realistic coronal hole structure \citep{Linker2017}. The PFSS estimates shown here do exhibit some limited time evolution over these two years of data but the fluctuations are in general much less than the difference (1) between different PFSS models (source surface height and magnetogram source) and (2) the difference between the model values and the in situ values. Overall, we observe that the open flux measured by PSP for the time interval from 2018-2020 (deep solar minimum) remains too high to be explained by standard PFSS estimates.

\section{Discussion}
\label{sec:disc}

\begin{table*}[h!]
  \begin{center}
    \caption{Open Flux and Heliospheric Flux measurements estimated from the literature. We note for results quoted in SI units of Wb we make the conversion to `field strength at 1AU' using $1 \text{nT AU}^2 = 4\pi (1.4955978707 \times 10^{11})^2 10^{-9} \text{Wb} \approx 2.81\times 10^{14} \text{Wb}$
    }
    \label{tab:other-measurements}
    \begin{tabular}{|p{0.18\textwidth}|p{0.05\textwidth}|p{0.2\textwidth}|p{0.2\textwidth}|p{0.26\textwidth}|}
      \hline
      \textbf{Reference} & \textbf{Epoch} & \textbf{$\Phi_{open}$ (nT AU$^2$)} & \textbf{$\Phi_H$ (nT AU$^2$)} & \textbf{Comments}\\
    \hline
    Present Study & 2018-2020 & 1.2 - 1.8 & $2.5^{+0.3}_{-0.6}$ & Parker Solar Probe, Figure \ref{fig:bulk_measurement} \\
    \hline
    \citet{Smith2011} & 2006-2007 && 2-2.5 & Ulysses, estimated from Fig. 4\\
    \hline
    \citet{Erdos2012} & Various & & $\sim$3 (solar min) & Ulysses (1990-2008), ACE (1998-2009), Helios (1975-1980), Conclusion 1 - 1997/1998 solar min \\
    \hline
    \citet{Erdos2014} & 1990- & & 3.5-6.5 (1991-1992), & Ulysses, OMNI, estimated from \\
    &2010&& 1.8-2.2 (1997-1998) & Fig 9. \\
    &&& 3-5 (2003-2004) & \\
    &&&  0.8-1.5 (2009-2010) & \\
    \hline
    \citet{Linker2017} & 2010 & 0.73 $\pm$ 0.1 (PFSS 2.5 $R_\odot$)   & 1.7-2.2 & Various Magnetograms, OMNI,\\
    & & 0.9 $\pm$ 0.1 (PFSS 2.0 $R_\odot$)  & &  Mean and SD computed from  \\
    & & 1.3 $\pm$ 0.1 (MHD) & & Table 1 \\
    \hline
    \citet{Owens2017} & 1999- & & 3.1-3,3 (2003-2004)  & ACE, estimated from Fig 5 (red  \\
    (Strahl method) &2011&& 1.4-1.6 (2009-2010) & curve) \\
    \hline
    \citet{Owens2017} & 1999- && ~2.8 (2003-2004) & ACE, estimated from Fig 5 (blue   \\
    (Kinematic Method) &2011&&  1.0-1.2 (2009-2010) & curve)  \\
    \hline
    \citet{Wallace2019} & 1990- & 2.5-4.6 (1991-1992) & 3.9-5.0 (1991-1992)  & NSO,ADAPT,OMNI, estimated  \\
    &2014& 1.8-2.1 (1997-1998) & 1.4-2.1 (1997-1998) & from Fig. 6\\
    && 1.8-2.5 (2003-2004) & 3.2-3.9 (2003-2004)  &    \\
    && 0.7-1.1 (2009-2010) & 1.1-1.2 (2009-2010) & \\
    \hline
    \end{tabular}
  \end{center}
\end{table*}

We examined the first five orbits of magnetic field data measured by Parker Solar Probe to measure the heliospheric magnetic flux ($\Phi_H$) down to 0.13AU and compared it to simultaneous measurements at 1AU, and estimates of the open magnetic flux escaping the corona as predicted by potential field source surface models. 

In this work, we use either 1 min averages or higher cadence data for which the distributions of data are not affected by the cancellation effect that appears with longer pre-averaging timescales.  \citet{Owens2017} showed that for 1AU data, an appropriate choice of a pre-averaging timescale (enacted on Cartesian components of the measured HMF vector field) has the same effect on the heliospheric flux measurement as using a more physically motivated method, such as accounting for local flux inversions. As shown in Appendix \ref{sec:pre-av}, pre-averaging even over one hour intervals of data strongly distorts the Cartesian components of the HMF vector and so, using any estimation method from such distributions is no longer a physical measurement of the actual magnetic field; and thus the `correct choice' of the pre-averaging timescale, while potentially useful \citep[e.g.  ][]{Wallace2019},  is more of a calibration technique which cannot be determined a priori. Moreover, it cannot be uniformly applied to data at different heliocentric distances as the character of the fluctuations changes. In this work, we utilise distributions of minimally pre-averaged data upon which our estimation methods (mean,mode, and Parker spiral method) are performed.  Furthermore, our implementation of the Parker spiral method (PSM) uses a high time-resolution polar representation of the HMF vector, which is much more weakly dependent on a particular pre-averaging timescale (appendix \ref{sec:pre-av}).

We show that the inclination of the Parker spiral and nature of vector fluctuations strongly affects the 1D distribution of measurements of $B_R R^2$ and, thus, can affect estimation methods that operate directly on the 1D distributions (the mean and mode). We also show (see Section \ref{sec:measured_flucs}) that the Parker spiral angle and fluctuation characteristics vary significantly with radius in the inner heliosphere and thus the specific distortion to the mean and mode also varies with radius. Under the assumption that the vector fluctuations are evenly distributed about a central value, these effects are mitigated with the PSM by computing the most probable 2D field vector in the R-T plane, and then projecting this into the radial direction, as was first suggested by \citet{Erdos2012}.  The applicability of the assumption of evenly distributed fluctuations was assessed in our computations of distribution skew in Figure \ref{fig:parker_spiral}, where we saw an increasing tendency towards skewed angular distributions close to 1AU (R > 0.8 AU). Thus, the PSM is likely less robust in this region as compared to closer distances to the sun. In addition,  non-fluctuation departures from the Parker spiral direction, for example magnetic clouds or field-line distortions near co-rotating interacting regions, may play a role and be a stronger influence further from the sun due to the increasingly inclined mean field and the slower orbital velocity of PSP meaning more such events are measured at these radii. Such distinct topological features would likely require a more detailed analysis such as through examining characteristics of the strahl \citep{Owens2017,Macneil2020} to remove these measurements from the population of measurements.

As observed by \citet{Smith2011} based on Ulysses data, we find that relative to the PSM, the effect of evenly distributed fluctuations which lie on a sphere is to systematically increase the value of the mean and mode with respect to a constant mean 2D vector as the Parker spiral grows more inclined, suggesting typical averaging methods which produce a growing estimate of $\Phi_H$ with radius are at least partially due to data processing artefacts. 

Such an enhancement in $\Phi_H$ with increasing heliocentric distance is indeed observed in the PSP and 1AU data. However, this enhancement is actually observed using the Parker spiral method, as well as the mean and the mode. This suggests either the Parker spiral method is also biased at 1AU, or the enhancement is at least partially physical. As noted above, however, the PSM may be less applicable near 1AU due to the less idealised nature of the angular fluctuations about a central value, which may explain why different behaviours were seen near 1AU for the positive polarity and negative polarity data in Figure \ref{fig:flux_vs_radius}. Locally inverted flux \citep{Owens2017} provides a plausible physical basis for such an enhancement and PSP data shows that the fraction of inverted flux grows with radius in agreement with Helios results, \citep[][]{Macneil2020}, although a simple estimate of the effect overestimates the observed discrepancy. In either case, the robustness of the PSP measurements is improved closer to the Sun. 

The inverted flux method described by \citet{Owens2017} remains an avenue to improve the estimates discussed in the paper especially with regard to quantifying more accurately the possible overestimation of the open flux further from the Sun. As remarked in Section \ref{sec:2d_vs_r}, data coverage of the electron heat flux from the PSP Solar Wind Electrons Alphas and Protons \citep[SWEAP; ][]{Kasper2016} instrument are less continuous than that of the magnetic field with FIELDS \citep{Bale2016} and, thus, such an estimate will only be possible for a subset of the time intervals investigated here (as an example, the outbound phase of encounter 3 will be missing). Nonetheless, given that near 1AU, all three methods considered in this work have issues, such a cross-check will be an important part of future work. We do expect, in any case, that such an investigation should conclude that the impact of inverted flux during PSP's closest approach to the Sun should remain minimal.

In Section \ref{sec:flux_vs_loc}, we computed measurements of $\Phi_H$ across three dimensions in the heliosphere to the extent permitted by the PSP and 1AU orbits. Compared against latitude, longitude, and radius (aside from the above inflection near 1AU), PSP data shows the heliospheric flux is well approximated as a conserved quantity throughout the inner heliosphere down to 0.13AU within at least +/- four degrees of the solar equatorial plane. Making a linear fit to the better resolved negative polarity PSM measurements (blue data points, top-left panel, Figure \ref{fig:flux_vs_loc}) suggests the open flux measured by PSP in the inner heliosphere varies by at most 0.1 nT AU$^2$ per AU.  A corollary to this observation is that the PSP data remains consistent with the Ulysses result \citep{Smith1995,Smith2003} that the open flux is uniformly distributed in latitude. The continuous applicability of the $1/R^2$ scaling implies the latitudinal reorganisation process which the coronal field undergoes as it expands into the heliosphere must be fully completed well within PSP's closest approach of 28 $R_\odot$. This observation is important to establish so that Equation \ref{eqn:hmf_method} can be used for all these PSP measurements and argue that measurements of $B_R R^2$ indeed constitute a direct measurement of the heliospheric flux, $\Phi_H$. We also note the important caveat that given our observation that fraction of inverted flux increases with radius (Figure \ref{fig:switchback_fraction}), this possible contribution to the flux should be quantified through methods such as described in \citet{Owens2017} before a definitive confirmation can be made that the the flat behaviour out to 1AU is continuous. In particular, this is necessary for determining at what radius inverted flux becomes a negligible contribution (which we infer from Figure \ref{fig:flux_vs_radius} to be approximately 0.8 AU).

Finally, in Section \ref{sec:flux_vs_time}, we compared the best estimate of $\Phi_H$ as a function of time to estimates from PFSS models. Both from the point of view of decaying relative amplitude of fluctuations, the fraction of field which is locally inverted, and the empirical observation that the same value was returned to on three consecutive perihelia at different longitudes, we conclude that the PSP measurements of the heliospheric flux at perihelion represent the most direct measurements to date of the total magnetic flux escaping the corona and, as such, impose a strong global constraint on coronal models.

This constraining value obtained with the PSM ($2.5^{+0.3}_{-0.6}$ nT AU$^2$) remains significantly higher that that predicted by standard PFSS models for this time interval (1.2-1.8 nT AU$^2$). As argued above, since PSP measurements are likely to be pristine measurements with little 'excess flux' effects as compared to 1AU, this provides a strong constraint and confirmation of a fundamental mismatch between current coronal models and the known flux in the inner heliosphere \citep[the 'open flux problem'][]{Linker2017}. 

In addition, we note the large discrepancy between the different magnetograms (e.g. the 2.0$R_\odot$ HMI model almost exactly matches the open flux estimate of the 2.5$R_\odot$ ADAPT-GONG model). This magnetogram-magnetogram disagreement is a well known issue with measurements of the photospheric field \citep{Riley2014, Wallace2019}, and in the absence of an independent way to calibrate these maps, correction factors have been proposed and evaluated by comparing to in situ measurements \citep[e.g.  ][]{Riley2007}. However, \citet{Wallace2019} establish, at minimum, the correction factors must be time-dependent (and, in particular, solar cycle-dependent) and conclude it is most likely that there are a number of reasons why the coronal model and in situ measurements disagree. One possible explanation is the lack of time dependence in PFSS and most MHD models which excludes contributions to $\Phi_H$ from transient disturbances, such as coronal mass ejections (CMEs), which can carry previously closed flux into the heliosphere \citep[e.g.  ][]{Owens2006}. 

We close our discussion by placing our results in context with a comparison to the actual values of open and heliospheric flux which have previously been reported in the literature. An important caveat here is that it has been well established \citep[e.g.  ][]{Wang2000,Erdos2014,Owens2017,Wallace2019} that the open flux varies with solar activity, solar cycle phase, and from one cycle to another. Thus, given the literature considered here is from different epochs, an exact agreement in measurement values is not to be expected. In Table \ref{tab:other-measurements}, we summarise estimates based on previous work. We note the values quoted are mostly estimated from graphical representations of data and so, they are not extremely precise. We estimate a representative range of values and focus on the years 1991-1992 and 2003-2004 for solar maxima, and 1997-1998 and 2009-2010 for solar minima. While these dates do not precisely match the true inflections of the solar cycle, among the literature studied, they are consistent with the times when heliospheric flux is observed to maximised or minimised.  The table lists the relevant reference, the epoch, the value of $\Phi_{open}$ if the paper uses coronal models, and $\Phi_H$ if the paper uses in situ determinations of the magnetic flux. For each reference, a note is made on the source of the data and the location in the relevant paper where we draw the estimate.

The literature used in our summary \citep{Smith2011,Erdos2012,Erdos2014,Linker2017,Owens2017,Wallace2019} spans epochs from 1990 through to 2014, encompassing solar minima in 1997 and 2008 and solar maxima in 1992, 2003, and 2014. The present study occurs in the next solar minimum (2018-2020). In general, the values determined (both for $\Phi_H$ and $\Phi_{open}$) are higher in this study compared to values during the 2008 solar minimum  and values determined by \citet{Linker2017} for data from 2010 (rising phase) and more similar to the 1997 solar minimum, but are less than all values determined for both solar maxima (1992,2003). The ratio of $\Phi_H/\Phi_{open}$ is similar to that of \citet{Linker2017} with values determined from a 2.5 $R_\odot$ PFSS model approximately half the value determined in situ. \citet{Wallace2019} found the vales to be closer around solar minimum 2008-2009 and their ratio ($\sim$0.7) is similar to the ratio of our 2.0 $R_\odot$ PFSS model to the lower bound of the in situ measurement. However, \citet{Wallace2019} also see almost identical PFSS and in situ measurements of the flux around solar minimum in 1997. 

Overall, the present work shows a consistent relationship between in situ and modelled open flux with prior works but also suggests the solar minimum studied in this work (2018-2019) had an overall higher flux baseline than the previous solar minimum (2009-2010) and, possibly, a worsening degree of the open flux problem as a result.

\section{Conclusions}
\label{sec:concl}

In this work we investigate the heliospheric flux content of the inner heliosphere with measurements that were taken closer to the corona than ever before, paying particular attention to the method used to estimate this quantity. We investigate the degree to which this quantity is conserved spatially, and compare it to the expected value of PFSS models. The main conclusions of our investigation are as follows: 

(1)  In computing the heliospheric magnetic flux, it is very important to recognise that 1D distributions of $B_R$ are, in fact, projections of a fluctuating vector. The balance of vector fluctuations, which show a measurable dependence on distance from the Sun, changes the $B_R$ distribution and, therefore, affects the interpretation of the 'background' value using typical statistical measures such as the mean or mode. Our implementation of the Parker spiral method \citep[PSM, ][]{Erdos2012,Erdos2014} is more robust than the mean or mode under the qualifying assumption of evenly distributed fluctuations of the vector about a central background value. Our implementation of the PSM is also improved on previous iterations by utilising an empirical Parker spiral angle and a high time-resolution (cadence $\leq$ 1 min) parameterisation of the heliospheric magnetic field (HMF) vector in polar coordinates, which is the most natural basis for analysing the vector fluctuations.

(2) As measured with the PSM, $B_R R^2$ is constant with longitude, latitude, and radius, at least for $0.13 < R < 0.8AU$ and for latitudes within four degrees of the solar equatorial plane. Thus, the PSP measurements are consistent with the \citet{Smith1995} and \citet{Smith2003} result that asserts that $B_R R^2$ is independent of latitude. In particular, the conservation with radius is evidence that the field is latitudinally isotropic at least down to 28 $R_\odot$ and so, it is well-motivated to assume Equation \ref{eqn:hmf_method_lon_int} holds and these single point measurements constitute a measurement of the heliospheric flux $\Phi_H$ for all probed radii.

(3) Except for intervals where PSP was very close to the HCS, the necessary assumptions of the PSM are well met during PSP's closest approaches to the Sun and, therefore, PSM measurements of $\Phi_H$ by PSP during these intervals constitute the most robust in situ estimate of the heliospheric flux in the inner heliosphere to date. Further to this, the much-reduced fluctuation amplitudes and the fraction of flux which is locally inverted suggests these measurements are insulated from physical excess flux; by following the Parker spiral method, they are robust to systematic data processing distortions. Thus, we argue that for these measurements, we can establish $\Phi_H \sim \Phi_{open}$ and this measurement constitutes a real constraint for the open flux that must be produced by coronal models. 

(4) The value measured for this period of time (October 2018 - July 2020 at solar minimum) is $\Phi_{H} = \Phi_{open} = 2.5^{+0.3}_{-0.6} nT AU^2,$ where the errors are the FWHM of the distributions of measurements close to the Sun (R < 0.3AU), as depicted in Figure \ref{fig:bulk_measurement}. This value appears higher than estimates for the previous solar minimum from around 2009-2010, but similar to prior estimates for the solar minimum from the late 1990s.

(5) This value is significantly larger than that implied by PFSS models driven by ADAPT maps over a reasonable range of source surface heights which produce estimates in the range 1.2 - 1.8 nT AU$^2$. From the results in \citet{Linker2017} reported in Table 1 above, as well as in Riley et al. (submitted to this issue), we see most MHD models do not close this gap. Therefore, we infer the resolution to the open flux problem is most likely to be found in new developments in heliospheric modelling or our knowledge of the photospheric field. Such improvements may come from Solar Orbiter \citep[SO; ][]{Muller2020} remote observations of the polar magnetic fields \citep{Solanki2020} or from the development of full time dependent coronal models which allow for processes such as interchange re-connection on a global scale \citep{Fisk1998,Fisk2020}, producing open field originating in non-coronal hole regions and which may be a more physical picture, as hinted in \citet{Boe2020}. Another idea suggested by \citet{Reville2020} was to include Alfv\'enic turbulence in an MHD simulation and this was seen to produce an envelope of fluctuations of the radial magnetic field which encompassed the amplitude of the PSP $B_R$ measurements from encounter 1.

Moving forwards, PSP will continue to dive deeper into the solar atmosphere in the coming years, eventually reaching a perihelion distance $< 10 R_\odot$. Given the current trends, the fluctuations in its magnetic field measurements are expected to continue to decay and the measurements discussed in this paper will likely become more and more accurate. There is some suggestion that the magnetic field could still be reorganising latitudinally as far out as $10 R_\odot$ \citep{Reville2017} and so we might expect that in its final orbit, PSP may start to detect a divergence from the $1/R^2$ trend observed in the radial field component so far, indicating a breakdown of the latitudinal isotropy which allows this powerful single point inference of the open magnetic flux. At the same time, such a measurement would be hugely exciting as it would constitute for the first time a direct measurement of the Sun's underlying dipole moment which, aside from being a fundamental quantity of interest, may also be able to address the calibration uncertainty in photospheric maps. We note that PSP's direct latitudinal sampling will remain confined to within four degrees of the solar equatorial plane as it continues to get closer. However, since the Solar Orbiter has now joined the PSP in the inner heliosphere, it will allow multi-point and out of the ecliptic in situ measurements \citep{Horbury2020b} to extend these results and provide even better and more direct constraints on the homogeneity of the heliospheric flux in the inner heliosphere. This includes directly sampling the solar wind eventually as far as 30 degrees from the solar equatorial plane, which will be important in robustly confirming conclusion (2) above.

\begin{acknowledgements}
Parker Solar Probe was designed, built, and is now operated by the Johns Hopkins Applied Physics Laboratory as part of NASA's Living with a Star (LWS) program (contract NNN06AA01C). Support from the LWS management and technical team has played a critical role in the success of the Parker Solar Probe mission. The FIELDS experiment on Parker Solar Probe spacecraft were designed and developed under NASA contract NNN06AA01C. The authors acknowledge the extraordinary contributions of the Parker Solar Probe mission operations and spacecraft engineering teams at the Johns Hopkins University Applied Physics Laboratory. S.D.B. acknowledges the support of the Leverhulme Trust Visiting Professorship program. S.T.B. was supported by the Mary W. Jackson NASA Headquarters under the NASA Earth and Space Science Fellowship Program Grant 80NSSC18K1201. This work utilises data obtained by the Global Oscillation Network Group (GONG) Program, managed by the National Solar Observatory, which is operated by AURA, Inc. under a cooperative agreement with the National Science Foundation. The data were acquired by instruments operated by the Big Bear Solar Observatory, High Altitude Observatory, Learmonth Solar Observatory, Udaipur Solar Observatory, Instituto de Astrofísica de Canarias, and Cerro Tololo Interamerican Observatory. This work utilises data produced collaboratively between Air Force Research Laboratory (AFRL) \& the National Solar Observatory (NSO). The ADAPT model development is supported by AFRL. The input data utilised by ADAPT is obtained by NSO/NISP (NSO Integrated Synoptic Program). This research made use of HelioPy, a community-developed Python package for space physics \cite{Stansby2019b}. This research has made use of SunPy v2.0.1 \citep{Mumford2019}, an open-source and free community-developed solar data analysis Python package \citep{Sunpy2015}.
\end{acknowledgements}

%
%

\bibliographystyle{aa}
\bibliography{OpenFluxPSP.bib}

\begin{appendix}
\onecolumn
\section{Justification of removing normal fluctuations}
\label{sec:appendix:normal_flucs}

\begin{figure*}[h!]
    \centering
    \includegraphics[width=\textwidth]{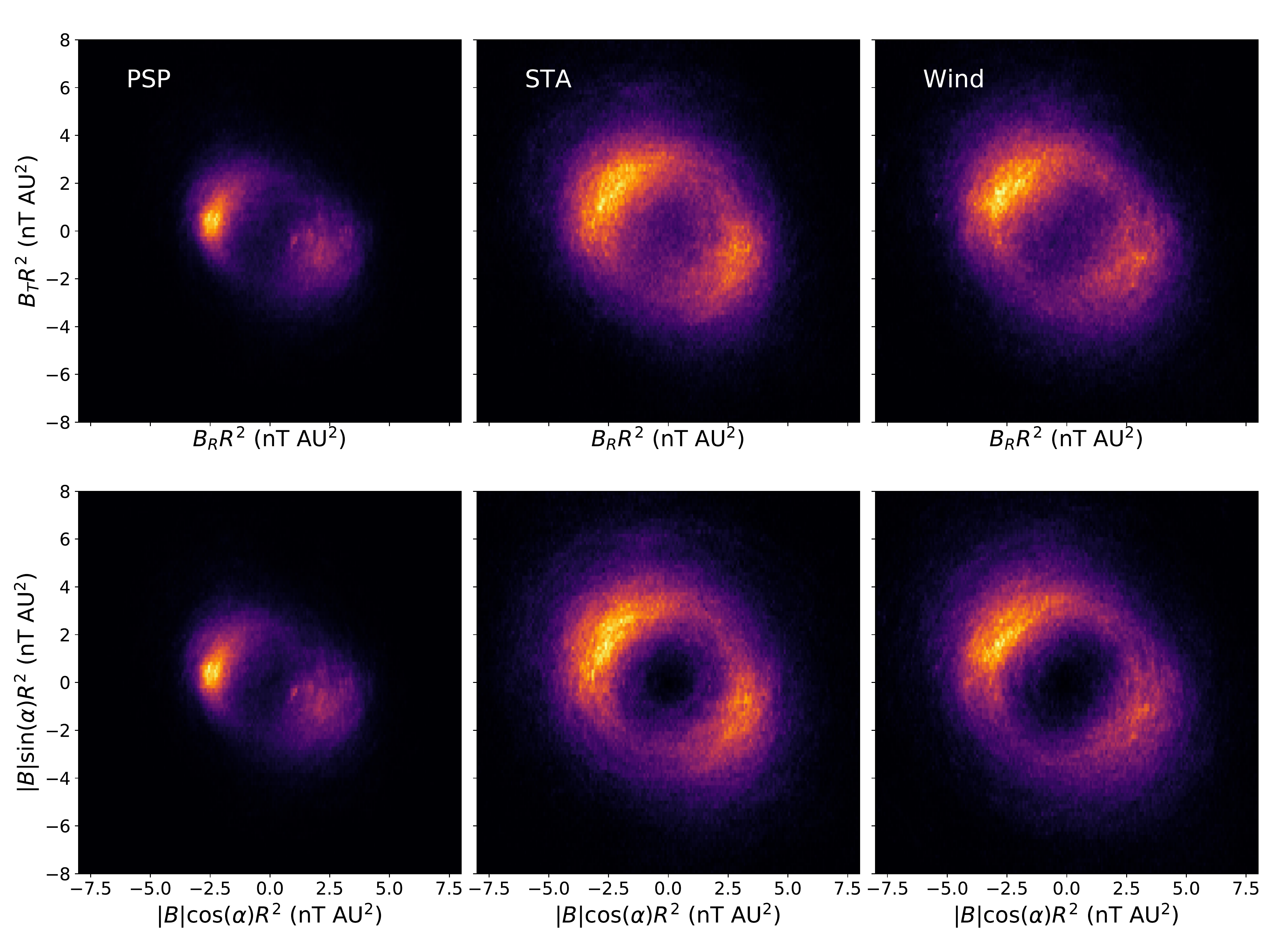}
    \caption{Justification of the treatment of the normal field component. Each panel shows a 2D distribution of field values across the full data set (see Figure \ref{fig:synoptic}). The top row shows the distribution of 1 min average values of $B_T R^2$ versus $B_R R^2$. The bottom row shows on the same axes and colour scale, the distribution formed by rotating the normal component into the R-T plane via the substitution $B_R = |B| \cos (\alpha)$, $B_T = |B| \sin (\alpha)$. The three columns show PSP, STA, and Wind data, respectively.
}
    \label{fig:appendix:normal_flucs}
\end{figure*}

In this paper, we treated the fluctuations as 2D (R-T plane) and assumed we can replace $B_R$ and $B_T$ with a parameterisation using the field magnitude and R-T clock angle. This is justified because the normal component generally fluctuates normally about the R-T plane, and these fluctuations are predominantly rotational (magnitude conserving). In addition, normal and tangential fluctuations are typically uncorrelated, and thus suppressing them in the vector magnitude-preserving method chosen here does not effect the distribution of vectors in the R-T plane. Figure \ref{fig:appendix:normal_flucs} shows the effect on the 2D distribution of making this substitution, with the top row of panels showing the 2D distribution of the raw $B_R R^2$ and $B_T R^2$ measurements for PSP, STEREO A, and Wind, respectively.  We note PSP data at all radii is here binned together. The bottom row shows the same distribution but with the normal component corrected for in the vector magnitude.

We see the effect at PSP is negligible, indicating the population of normal fluctuations is sub-dominant to fluctuations in the R-T plane. The effect at 1AU is quite striking. We see that just by taking the raw $B_R$ and $B_T$ values in a 2D distribution, a large population of data points exist near the origin. These data are actually just a projection of the normal component onto the R-T plane. Therefore, this population is nonphysical: from the 2D distribution, we would conclude there is a large population of near-zero field magnitude, whereas, in fact, from the bottom row of Figure \ref{fig:appendix:normal_flucs}, we see by making this normal correction, this population vanishes and we get a more accurate measure of the RT distribution, especially the magnitude. Thus, we conclude it is a robust and useful transformation to correct the R-T components of the field by the normal component and use this data throughout the paper.

\section{Synthetic versus real distributions}
\label{sec:appendix:synth_vs_real}

\begin{figure*}[h!]
    \centering
    \includegraphics[width=\textwidth]{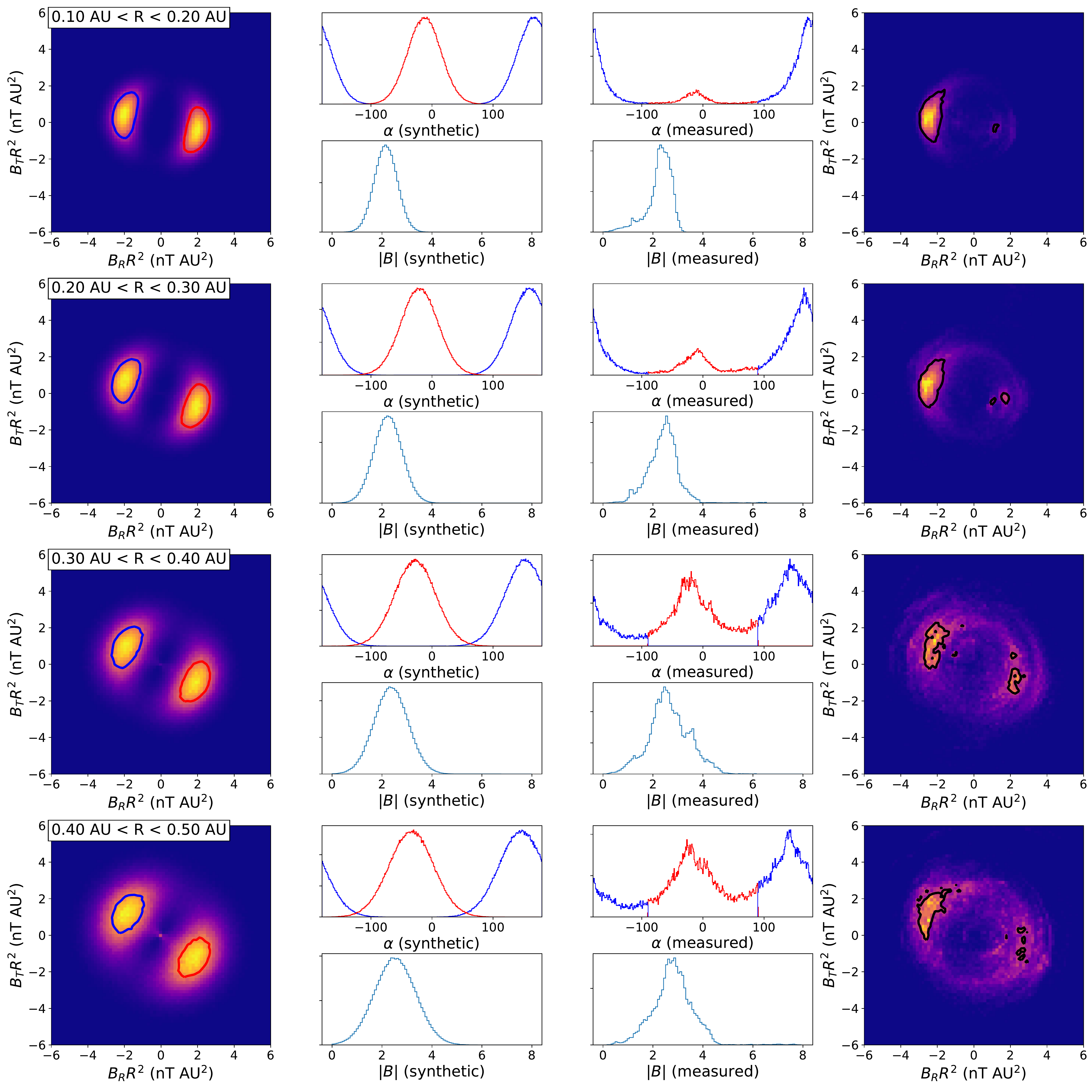}
    \caption{Synthetic and measured flux distributions as a function of radius (0.1AU-0.5AU). The left-hand column shows the 2D synthetic distributions of $B_T R^2$ versus $B_R R^2$, the rightmost column shows the corresponding measured distribution. For both 2D histograms, a contour shows the 90th percentile of the data. The middle columns consist of corresponding 1D distributions of clock angle (top panel for each radial bin) and field vector magnitude (x $R^2$) (bottom panel for each radial bin). For the clock angle, red (blue) curves represent anti-sunward (sunward) sector populations. }
    \label{fig:appendix:synth_vs_real_close}
\end{figure*}

\begin{figure*}[h!]
    \centering
    \includegraphics[width=\textwidth]{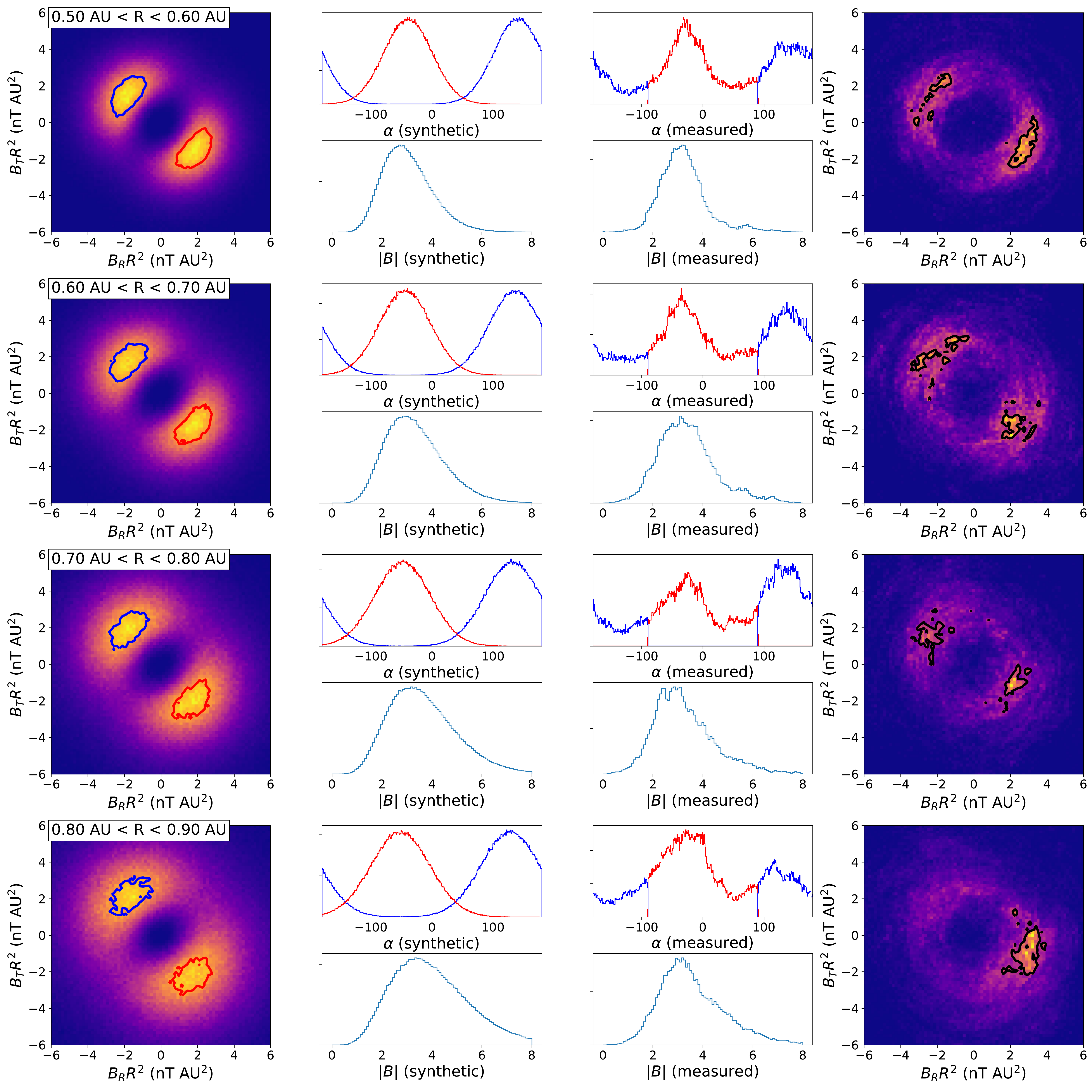}
    \caption{Continuation of Figure \ref{fig:appendix:synth_vs_real_close} : Synthetic and Measured Flux distributions as a function of radius (0.1AU-0.5AU).}
    \label{fig:appendix:synth_vs_real_far}
\end{figure*}

In Appendix \ref{sec:appendix:synth_vs_real}, we compute estimates of $B_R R^2$ for data binned by radius (meaning from multiple orbits) and compared them to estimates from derived synthetic data. In Figures \ref{fig:appendix:synth_vs_real_close} and \ref{fig:appendix:synth_vs_real_far}, we display the corresponding 2D distributions and 1D distributions in magnitude and clock angle for radial bins from 0.1 to 1.0 AU. We note that these bins are slightly wider than what we describe in Section \ref{sec:measured_trends}, for the sake of brevity. These plots show how the large scale variation and growth of fluctuations and Parker spiral background is captured by the synthetic data and the generic 2D shape compares well to the raw data. We also see that the 1D distributions we use to approximate the magnitude and the clock angle do have limitations. In particular, the real distributions are spikier and in some cases appear to show different streams merged together, while the synthetic data assumes one smooth population. The distributions of clock angle are generally formed more like a triangular-shaped distribution compared to the approximated Gaussian distributions. We also see a systematic skew in the the magnitude with higher radii. For cases where the skew is large enough that a normal distribution with the same standard deviation would predict negative values of |B|, we allow the synthetic distribution to have non-zero skew. As noted in the main text, in these cases we continue to interpret the distribution peak (rather than the mean) as the `central value' about which fluctuations occur.

\section{Pre-averaging and vector representations}
\label{sec:pre-av}

\begin{figure}[h!]
    \centering
    \includegraphics[width=0.7\textwidth]{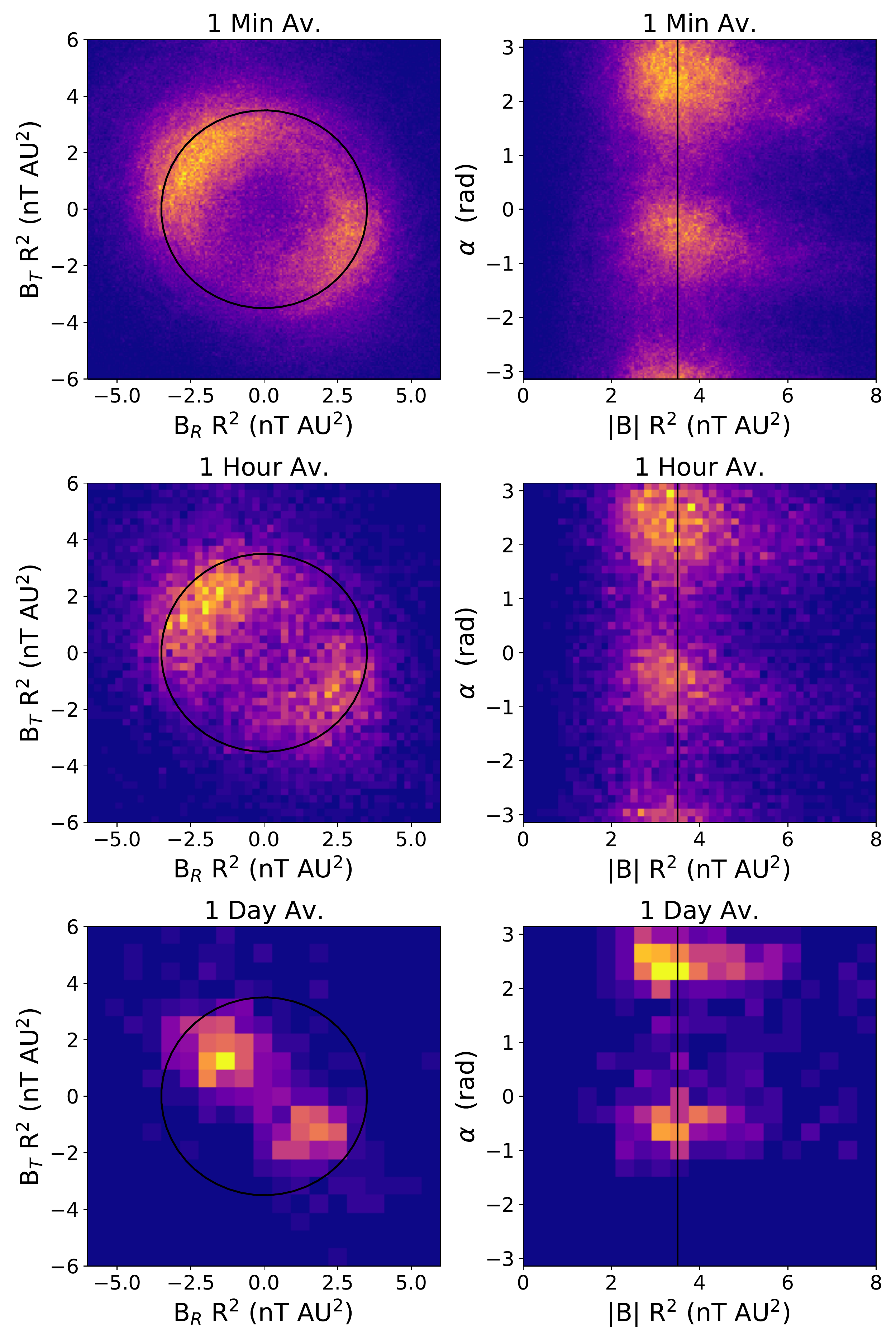}
    \caption{Demonstration of the effect of pre-averaging HMF vector data on 2D data distributions under Cartesian and spherical representations. Each panel shows a 2D histogram of data from STEREO A for the time interval considered in this work. From top to bottom, successively more aggressive pre-averaging is applied to the data set with panels showing one minute, one hour, and one day averages. The left hand panel shows the Cartesian components $B_R R^2$ versus $B_T R^2$, while the right hand column shows the spherical representation (clock angle versus magnitude). The histogram resolution is decreased with successive averages as the number of data points reduces. A black circle in the left hand column and vertical line in the right hand column shows a curve of $|B|R^2 = 3$nT AU$^2$ as a guide and point of comparison between the different averages and shows the averaging transformation is more magnitude preserving when applied to the polar representation as compared to the Cartesian representation.
    }
    \label{fig:appendix:pre-av}
\end{figure}

As discussed in Section \ref{sec:2d_vs_r}, pre-averaging raw spacecraft data to a lower cadence can have an impact on the data distributions in a non-trivial way. For vector data, the coordinate representation can have an impact on the effect of pre-averaging. In Figure \ref{fig:appendix:pre-av}, we demonstrate this with the STEREO A 1 min averaged base data product which was introduced in Figure \ref{fig:synoptic}. From top to bottom, the panels show one-minute, one-hour, and one-day averages of the vector magnetic field. The left-hand column shows the data parameterised in Cartesian coordinates (R and T), while the right-hand column shows the data parameterised in polar coordinates by the vector magnitude and clock angle ($\alpha$ in the main text). The top panels show exactly the same data ($B_R = |B|\cos \alpha$, $B_T = |B| \sin \alpha$. While in Cartesian distribution it fills out an annular region of parameter space, in polar coordinates, the region is more like two Gaussian ellipses. In particular, the major and minor axes of these ellipses are aligned with the axes in this case, suggesting fluctuations in |B| and $\alpha$ have low correlation - their fluctuations are independent. The further panels are generated by producing averages from the above 1 min averaged data, meaning that for the polar coordinate plots, the time series $|B|(t)$ and $\alpha(t)$ are the quantities that are averaged. For both cases, the resolution of the histograms worsens with larger averaging due to the lower statistics.

In both columns, a consistent contour of $|B|R^2 = 3 $nT AU$^2$ is shown as a black curve. This is a circle in Cartesian and a vertical line in polar coordinates. For the Cartesian coordinates, as mentioned in the main text, the distribution is strongly distorted, especially noticeable in the one day averages. While the raw data is quite tightly confined to an annulus but spread quite widely around that annulus, averaging reduces the spread around the annulus but importantly also causes the data to migrate towards the origin, effectively reducing the field magnitude of the distribution. While the averaging out of the angular fluctuations is arguably a useful effect, the magnitude effect is an artificial distortion.

In polar coordinates, on the other hand, the only perceptible change of the distribution is the worsening resolution. Since the data are quite well distributed around a mean and the distribution is aligned with the $|B|$ and $\alpha$ axes, when successive averages of samples from this distribution are taken, the resulting means are essentially in the same distribution. As mentioned above, an equivalent inference is that the fluctuations in $|B|$ and $\alpha$ are independent or uncorrelated. The 3nT AU$^2$ contour cuts the distribution at approximately the same place. Thus, using this polar representation of the vector is useful since it produces vector time series whose distribution is much less affected by time averaging as compared to the Cartesian representation.

\section{Schematic of a local field-line inversion}
\label{sec:appendix:sb_schem}

\begin{figure}[h!]
    \centering
    \includegraphics[width=0.5\textwidth]{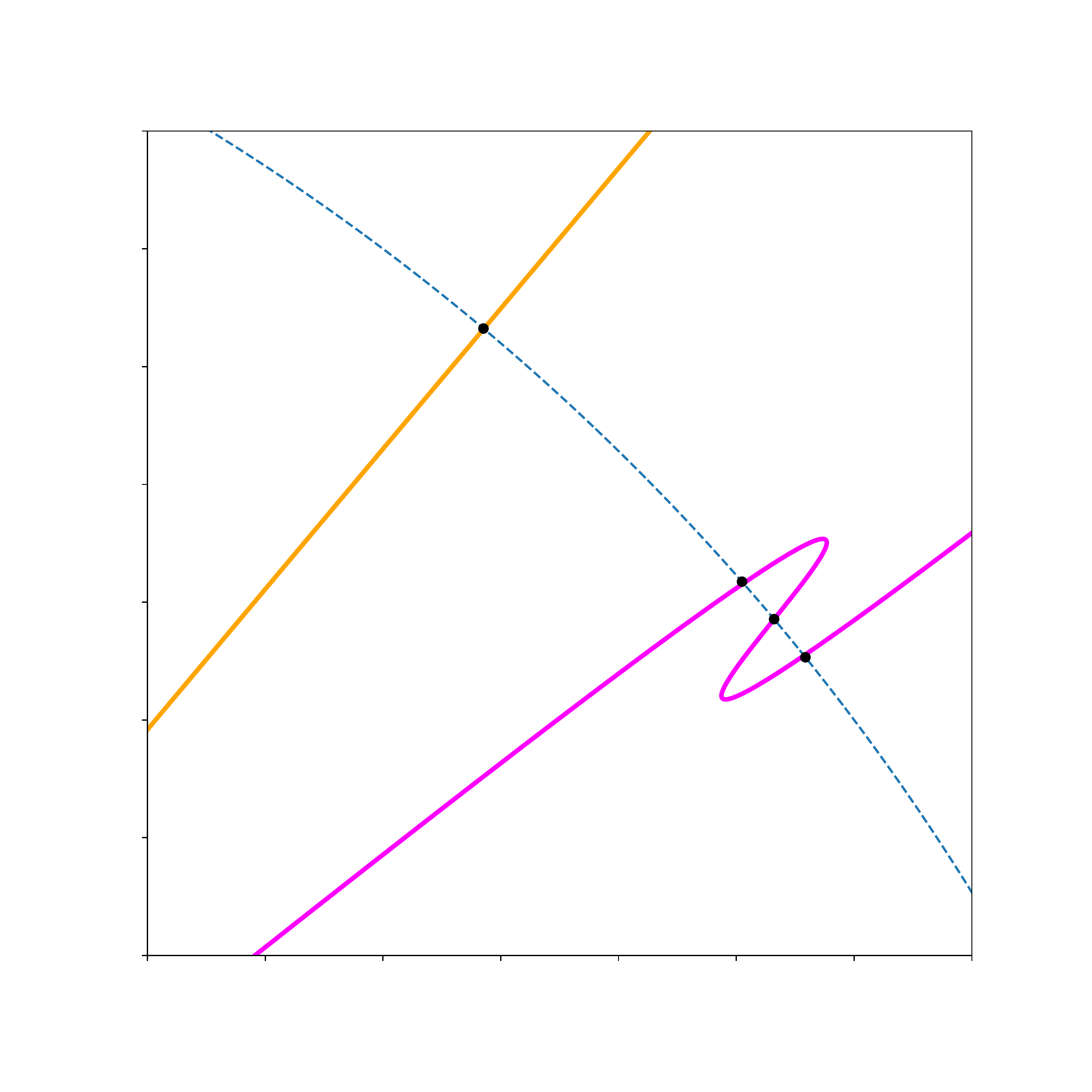}
    \caption{Schematic of of a local field inversion and it’s impact on the flux. A standard, non-inverted field-line (orange) and a field line which folds back on itself (magenta). The inverted field line intersects the spherical surface (dotted line) three times and therefore has a three times greater contribution to the flux at this radius compared to it’s contribution in escaping the corona.}
    \label{fig:appendix:sb_schem}
\end{figure}

In Section \ref{sec:excess_flux}, we discussed the possible contribution of excess flux due to local topological inversions of the magnetic field, as per \cite{Owens2017}. Here we illustrate schematically the statement that `inverted field lines contribute three times the flux as non-inverted field lines'. In this schematic, we show a surface of constant radius (dotted black line) and a non-inverted (orange), and inverted (magenta) field line. Black scatter points indicate where these curves intersect with the radial surface. As we can see, the inverted field line intersects the surface three times, compared to the orange curves single intersection. Thus, when we conserve flux by tracing field lines from the corona out to this radius, the orange field line will contribute the same flux at the corona and this outer radius, but the inverted line will contribute three times as much. This means when contributions are all summed up at the two different radii, there will be a larger flux and, therefore, a larger value for $B_R R^2$ at the outer radius.

\end{appendix}

\end{document}